%Paper: hep-th/9509170
%From: Clifford Johnson <cvj@itp.ucsb.edu>
%Date: Thu, 28 Sep 95 20:37:09 PDT
%Date (revised): Tue, 10 Oct 95 22:30:24 PDT

%%%%%%%%%%%%%%%%%%%%%%%%%%%%% %%%%%%%%%%%%%%%%%%%%%%%%%%%%%%%%%
\input harvmac
%-------------------------------Journals----------------------------
\def\npb#1#2#3{{ Nucl. Phys. }{\bf B#1} (#2) #3}
\def\plb#1#2#3{{ Phys. Lett. }{\bf B#1} (#2) #3}

\def\prl#1#2#3{{ Phys. Rev. Lett. }{\bf #1} (#2) #3}

\def\cmp#1#2#3{{ Commun. Math. Phys. }{\bf #1} (#2) #3}

%----------------------------Abbreviations---------------------------

\def\bar{\overline}
\def\e#1{{\rm e}^{\textstyle#1}}

\def\darr#1{\raise1.8ex\hbox{$\leftrightarrow$}\mkern-19.8mu #1}
\def\half{{\textstyle{1\over2}}} %puts a small half in a displayed eqn
\def\roughly#1{\ \lower1.5ex\hbox{$\sim$}\mkern-22.8mu #1\,}

%%%%%%%%%%%%%%%%%%%%%%%%%%%%%%%%%%%%%%%%%%%%%%%%%%%%%%%%%%%%%%
\hyphenation{di-men-sion di-men-sion-al di-men-sion-al-ly}

\parindent=10pt
\parskip=5pt

\def\Tr{{\rm Tr}}
\def\det{{\rm det}}

\def\wzw{Wess--Zumino--Witten}
\def\Az{A_z}
\def\Azb{A_{\bar{z}}}

\def\d{\partial_z}
\def\db{\partial_{\bar{z}}}

\def\IR{{\hbox{{\rm I}\kern-.2em\hbox{\rm R}}}}
\def\IB{{\hbox{{\rm I}\kern-.2em\hbox{\rm B}}}}
\def\IN{{\hbox{{\rm I}\kern-.2em\hbox{\rm N}}}}
\def\IC{{\hbox{{\rm I}\kern-.6em\hbox{\bf C}}}}
\def\IZ{{\hbox{{\rm Z}\kern-.4em\hbox{\rm Z}}}}
\def\half{{{1\over 2}}}
\def\X{{\cal X}}

\def\tm{{\tilde{m}}}
\def\tn{{\tilde{n}}}
\def\bc{{\bar{c}}}

\def\bL{{\bar{L}}}
\def\bC{{\bar{C}}}
\def\bth{{\bar{\Theta}}}
\def\Sb{{\sum_{b=0}^{Q^2-2}}}

\noblackbox
\def\WZW{Wess--Zumino--Witten}
\def\D{{\cal D}}

\font\bigtitlerm=cmr10 scaled \magstep3

\catcode`\@=11
\nopagenumbers\abstractfont\hsize=\hstitle
\null%\vskip-40pt
\rightline{\vbox{\baselineskip12pt\hbox{NSF-ITP-95-117}
                                  \hbox{HUTP--95/A002, MIT--CTP--2463}
                                  \hbox{IASSNS-HEP-95/68, PUPT--1553}
                                   \hbox{hep-th/9509170}}}
\vfill\vskip.5cm
\vskip5pt
\centerline{\bigtitlerm Heterotic Coset Models and (0,2) String Vacua}
\vskip5pt
\abstractfont\vfill\pageno=0
\vskip.5cm
\centerline{\bf Per Berglund$^{a,*}$\footnote{}{\sl $^*$After 1st Jan. 1996:
Theory Division, CERN, 1121 Geneva 23 Switzerland.},
Clifford V. Johnson$^a$, Shamit Kachru$^b$,
Philippe Zaugg$^c$}                                \vskip0.5cm
\baselineskip=11pt
\centerline{\sl $^a$Institute For Theoretical Physics, University of
California,}
\centerline{\sl Santa Barbara CA 93106 USA}
\centerline{\sl berglund@itp.ucsb.edu, cvj@itp.ucsb.edu}\vskip4pt
\centerline{\sl $^b$Lyman Laboratory of Physics,  Harvard University}
\centerline{\sl Cambridge MA 02138 USA}
\centerline{\sl kachru@string.harvard.edu}\vskip4pt
\centerline{\sl $^c$Center for Theoretical Physics,
Massachusetts Institute of Technology,}
\centerline{\sl  Cambridge, MA 02139 USA}
\centerline{\sl zaugg@mitlns.mit.edu}
\vfill
\vskip0.2cm
%\vskip-0.3cm
\centerline{\bf Abstract}\vskip0.5cm
\vbox{\narrower\baselineskip=12pt\noindent
A Lagrangian definition of a large family of $(0,2)$ supersymmetric
conformal field theories may be made by an appropriate gauge invariant
combination of a gauged  Wess--Zumino--Witten model,
right--moving supersymmetry
fermions, and left--moving current algebra fermions.
Throughout this paper,  use is made of the interplay between
field theoretic and algebraic techniques (together with supersymmetry)
which is facilitated by such a definition.  These heterotic coset models
are thus studied in some detail, with
particular attention paid to the $(0,2)$ analogue of the
$N{=}2$ minimal models,
which coincide with the `monopole' theory of Giddings, Polchinski and
Strominger.
A  family of modular invariant partition
functions for these $(0,2)$ minimal models
  is presented. Some  examples of $N{=}1$
supersymmetric four dimensional
string theories with gauge groups  $E_6{\times} {\widetilde G}$
and $SO(10){\times} {\widetilde G}$
are presented, using these minimal models as building blocks. The factor
${\widetilde G}$
represents various enhanced symmetry groups made up of products of
$SU(2)$ and $U(1)$.}

\Date{September 1995}

%\draft

\baselineskip13pt

\newsec{Introduction}
As is by now common knowledge, the presence of $N=1$ supersymmetry in four
 dimensional spacetime results in $(0,2)$ supersymmetry on the heterotic string
worldsheet,
together with the condition that   states have odd integral right $U(1)$
charges\ref\susy{L. Dixon, in `Proceedings of the 1987 ICTP Summer Workshop in
High Energy Physics and Cosmology', edited by G. Furlan, et. al.}\ref\banks{T.
Banks, L. Dixon, D. Friedan and E. Martinec, \npb{299}{1988}{613}\semi
A. Sen, \npb{278}{1986}{289}.}.  The study of superstring  vacua with
$N=2$ worldsheet supersymmetry is quite a mature subject by now, but chiefly in
the specialized area of $(2,2)$ vacua. This situation is largely due to the
fact (observed in ref.\ref\instant{X. G. Wen and E. Witten,
\plb{166}{1986}{397}\semi M. Dine, N. Seiberg, X. G. Wen and E. Witten,
\npb{278}{1986}{769} and
\npb{289}{1987}{319}.}) that generically, $(0,2)$ sigma models appeared to
have dangerous
worldsheet instanton effects. It was observed\ref\resurrection{J. Distler,
\plb{188}{1987}{319}\semi
M. Cvetic, \prl{59}{1987}{2829}\semi
M. Dine and N. Seiberg, \npb{306}{1988}{137}\semi
J. Distler and B. Greene, \npb{304}{1988}{1}.}\ that $(0,2)$ models could
nevertheless be shown to exist, but  $(2,2)$ models were seen to be much
easier to define and study, given the techniques available at that time.

It is certainly time to try to close the gap opened in our  understanding
of $(2,2)$ versus
 $(0,2)$ vacua, if we are ever going to honestly claim that we
have some understanding of string  theory in generic backgrounds. Indeed, the
renaissance of $(0,2)$ models can already be said to have begun.
Recent results have breathed new life into the study of $(0,2)$ models,
primarily due to the invention of new techniques for defining and studying
$N=2$ models in general. The `linear $\sigma$--model' approach to $N=2$
models\ref\edlinear{E. Witten, \npb{403}{1993}{159}, hep-th/9301042.}\ allowed
for a completely
new approach to understanding $N=2$ string vacua, casting  Calabi--Yau and
Landau--Ginsburg formulations in the same framework and showing how $(0,2)$
models can arise simply as deformations of $(2,2)$ models. Continuing the
development of these techniques, the work presented in
ref.\ref\dkone{J. Distler and S. Kachru, \npb{413}{1994}{213},
hep-th/9309110.}\ defined a much
larger class of $(0,2)$ models (not only deformations of $(2,2)$'s),
and refs.\ref\dktwo{J.
Distler and S. Kachru, \npb{430}{1994}{13}, hep-th/9406090.}\ref\evaone{E.
Silverstein and
E. Witten, Nucl. Phys. {\bf B444} (1995) 161, hep-th/9503212.}\ref\evatwo{E.
Silverstein and E.
Witten, Phys. Lett. {\bf B328} (1994) 307, hep-th/9403054.}\ref\per{P.
Berglund, P. Candelas,
X. de la Ossa, E. Derrick, J. Distler, and
T. Hubsch, {\sl `On the Instanton Contributions to the Masses and Couplings of
$E_{6}$ Singlets,'} hep-th/9505164, to appear in Nucl. Phys. {\bf B}.}\
showed that the
conditions for conformal invariance in $(0,2)$ models defined in this way
 are likely to be satisfied by a
very large class of models.

So the situation suddenly looks much better for $(0,2)$ models, although we
have not yet attained the level of understanding which we have of $(2,2)$
models. Another step in this direction would be to have a large class of
exactly solvable conformally invariant models for use as a laboratory,
as is traditional in almost
any area of theoretical endeavour. The $N=2$ minimal models  and the rest of
the Kazama--Suzuki models have played this role in the realm of $(2,2)$ models
and have been of  importance in the understanding of
the moduli space of $(2,2)$ vacua. It would certainly be very pleasing to
have the analogous family of building blocks with which to continue studying
$(0,2)$ moduli space.

Generally, the search for modular invariant combinations of characters which
define  conformal field theories with such a highly non--diagonal structure
is a daunting task if there is  little physical guidance. However,
there have been a number of successful searches using
powerful algebraic techniques\ref\cftone{T. Gannon, Nucl.Phys.{\bf B402} (1993)
729, hep-th/9209042\semi T. Gannon and Q. Ho-Kim, Nucl. Phys.
{\bf B425} (1994) 319, hep-th/9402027.}\ref\cfttwo{R. Blumenhagen and
A. Wisskirchen, {\sl `Exactly Solvable (0,2) Supersymmetric String Vacua With
GUT Gauge Groups'}, preprint BONN-TH-95-11, IFP-507-UNC, hep-th/9506104.}.

In ref.\ref\gps{S. B. Giddings, J. Polchinski and A. Strominger,
Phys. Rev. {\bf D48}
(1993) 5784, hep-th/9305083.}, in a study of magnetically charged  four
dimensional black hole
solutions of string theory, a family of conformal field theories (CFT's) were
constructed as asymmetric orbifolds of affine $SU(2)$. These CFT's, when
specialized to  the heterotic string case, are examples of $(0,2)$ models and
therefore have applications other than as $U(1)$ monopole backgrounds, as was
noted in the discussion in ref.\gps. In refs.\ref\paperone{C. V. Johnson, Phys.
Rev. {\bf D50} (1994) 4032, hep-th/9403192.}\ref\papertwo{C. V. Johnson, Mod.
Phys. Lett. {\bf
A10} (1995) 549, hep-th/9409062.}\ref\paperthree{C. V. Johnson and R. C. Myers,
Phys. Rev. {\bf D52}  (1995)
2294,
  hep-th/9503027.} it was shown that this `monopole' theory was an example of a
larger class of $(0,2)$ CFT's (`Heterotic Coset Models') which may be
formulated directly
by Lagrangian methods. These methods were used to generalize the CFT
construction of magnetic black holes to dyonic black holes, Taub--NUT, and
Kerr--Taub--NUT solutions of Heterotic String theory. These heterotic coset
models
 are a  non--trivial
coupling  of
gauged \wzw\ models with world sheet fermions. It is immediately
apparent  that the models
 thus defined are very close in
spirit and structure to the familiar $N=2$ minimal models and the rest of the
Kazama--Suzuki models. In fact, those $(2,2)$ cases can be seen as special
cases (but {\sl not}, in general, deformations) of the $(0,2)$ models defined
there. In this way, we  now have a family of exactly
solvable $(0,2)$
building blocks analogous to the one known for the $(2,2)$ case.
This  allows us to search for  analogues of the Gepner
points\ref\gepner{D. Gepner, Phys. Lett. {\bf B199} (1987) 380;
Nucl. Phys. {\bf B296} (1988) 757.}\ in the moduli space of $(0,2)$
compactifications. Evidence for such points has recently been given in
ref.\cfttwo.

This paper continues the work of ref.\papertwo, by starting with the Lagrangian
definition of the models presented there and proceeding to study their
content. It is here that the power of field theory makes its presence felt, as
 a great deal can be written down quite readily about the partition
 function by using the field theory as a guide. Focusing on the direct analogue
 of the $N=2$ minimal models, which shall be referred to as the
`$(0,2)$ minimal models', we proceed to compute their elliptic genera in
 section~3, showing  (in the spirit of ref.\ref\edminimal{E. Witten, Int. J.
Mod. Phys. {\bf A9} (1994) 4783, hep-th/9304026.}, and following the
computation of ref.\ref\mansi{M. Henningson, Nucl. Phys. {\bf B413} (1994) 73,
hep-th/9307040.})
how the presence of supersymmetry coupled with a
Lagrangian definition allows a great amount of information about the
elliptic genera of the models
to be extracted readily. In section~4 much further progress is made, by using
more field theory techniques to motivate in great detail the form of the
complete partition function and then completing the computation,
checking modular invariance.
 Thus armed with a  store of $(0,2)$ modular
invariants we proceed in section~5 to study the massless
spectrum of a handful of
four dimensional string theories with $SO(10)\times {\widetilde G}$ and
$E_6\times {\widetilde G}$
gauge groups and $N=1$ spacetime
supersymmetry. The factor ${\widetilde G}$ is an enhanced gauge symmetry group
arising from the details of the internal conformal field theory, and
generically takes the form of a product of $SU(2)_6$ and $U(1)$ factors.
We conclude with a discussion in section~6.

\nref\novikov{S. P. Novikov, Ups. Mat. Nauk. {\bf 37} (1982) 3.}
\nref\edwzw{E. Witten, Commun Math Phys {\bf 92} (1984) 455.}

\newsec{Lagrangian definitions of Conformal Field Theories}
\subsec{(Super) Wess--Zumino--Witten Models}
The starting point for a Lagrangian definition of a conformal field theory for
the purposes of this paper is the \wzw\ (WZW) model\novikov\edwzw,
based on a compact  semi--simple Lie group $G$ at level $k$:
\eqn\iwzw{\eqalign{&I_{WZW}(g,k)=\cr&-{k\over4\pi}\int_{\Sigma} d^2z\,\,
\Tr[g^{-1}\d g\cdot g^{-1}\db
g]-{ik\over12\pi}\int_B d^3\sigma\,\,
\epsilon^{ijk}\Tr[g^{-1}\partial_ig\cdot g^{-1}\partial_jg\cdot g^{-1}
\partial_kg]}}
where the two dimensional surface $\Sigma=\partial B$ has coordinates
$(z,{\bar z})$ and $g\in G$.
This model has a `global' $G_L\times G_R$ symmetry $g\to g^{\phantom R}_L(z) g
 g_R^{-1}({\bar z})$
where $(g^{\phantom R}_L,g^{\phantom R}_R)\in
(G^{\phantom R}_L,G^{\phantom R}_R)$.
This results in the model's huge success as a
solvable system, due to the resulting affine Lie  algebra satisfied by the
currents\ref\kz{V. G. Knizhnik and
A. B. Zamolodchikov, Nucl. Phys. {\bf B247} (1984)
83.}\ref\gepwit{D. Gepner and E. Witten, Nucl. Phys. {\bf B278} (1986) 493.}.
Essentially, this model may be thought of as the conformal field theory of
a string propagating on a group manifold $G$\gepwit.

There exists a supersymmetric extension of this model\ref\rohm{R. Rohm, Phys.
Rev. {\bf D32} (1984) 2849.}. One way that
 this may be discovered is by using a superfield
construction of the supersymmetric WZW model.
For our starting point, we shall simply write a supersymmetric WZW by putting
in the free fermions (in the adjoint representation)
 immediately at the component
level\ref\ed{E. Witten, Nucl. Phys. {\bf B371} (1992) 191.}:
\eqn\quantum{I^{(1,1)}=I_{WZW}(g,{k})
+{i{k}\over4\pi}\int_\Sigma d^2z\,\,\Tr[\Psi_R\db\Psi_R+\Psi_L\d \Psi_L],}
noting that  supersymmetry is realized (on shell) simply as:
\eqn\susyone{\eqalign{
&\delta g=i\epsilon_Rg\Psi_R+i\epsilon_L\Psi_Lg\cr
&\delta \Psi_R=\epsilon_R(g^{-1}\d g-i\Psi_R\Psi_R)\cr
&\delta \Psi_L=\epsilon_L(\db gg^{-1}+i\Psi_L\Psi_L)\cr
}.}

\subsec{Lagrangians for $(1,1)$ Coset Models}
Coset models\ref\bardhal{K.
Bardakci and M. B. Halpern, Phys. Rev. {\bf  D3} (1971) 2493\semi M. B.
Halpern, Phys. Rev. {\bf D4} (1971) 2398.}\ref\gko{P. Goddard and D. Olive,
Nucl. Phys. {\bf B257} (1985)
226\semi P.
Goddard, A. Kent and D. Olive, Phys Lett {\bf B152} (1985) 88\semi
P. Goddard, A. Kent and D. Olive, Commun Math Phys {\bf 103} (1986) 105.}\
are algebraic constructions of
conformal field theories based on the current algebras of $G$ and a subgroup
$H$. They may be given a Lagrangian realization by the use of gauged WZW
models, where (naively) the picture of a string moving on a group manifold is
replaced by one of having the string restricted to moving on the
subspace given by the coset $G/H$. This is realized
consistently as a conformal field theory by
gauging away the unwanted degrees of freedom corresponding to movement
outside the coset,  constructing a gauge--invariant extension of the WZW model,
by introducing non--propagating 2D gauge fields $A_a$.
That this corresponds to the algebraic coset construction has been shown to a
great
extent over the last few years\ref\cosetgWZW{D. Karabali, Q--H. Park, H. J.
Schnitzer and Z. Yang, Phys. Lett. {\bf
B216} (1989) 307\semi D. Karabali and  H. J. Schnitzer, Nucl. Phys. {\bf B329}
(1990) 649--666.}\ref\gk{K. Gawedski and
A. Kupiainen, \npb{320}{1989}{625}.}\ref\hwangco{S. Hwang and H. Rhedin,
 Nucl. Phys. {\bf B406} (1993) 165, hep-th/9305174.}, using a variety of
methods.

The (naively) $N=1$ supersymmetric Lagrangian
for coset models was written down and studied
in ref.\ref\shnitzer{H. J. Schnitzer, Nucl. Phys. {\bf 324} (1989) 412.}.
In the component form, we may write the action as follows:
\def\DG{{g^{-1}\d g}}\def\DBG{{\db gg^{-1}}}
\eqn\SGWZW{\eqalign{I^{(2,2)}& =I_{WZW}(g)+I(g,A)+I_F(\Psi_L,\Psi_R,A)=\cr
&-{k\over4\pi}\int_{\Sigma} d^2z\,\, \Tr[g^{-1}\d g\cdot g^{-1}\db
g]\cr &-{ik\over12\pi}\int_B d^3\sigma\,\,
\epsilon^{ijk}\Tr[g^{-1}\partial_ig\cdot g^{-1}\partial_jg\cdot g^{-1}
\partial_kg]\cr
&+{k\over2\pi}\int_\Sigma d^2z\,\,\Tr[\Azb\DG-\Az\DBG+\Azb g^{-1}\Az
g-\Az\Azb]\cr
&+{ik\over4\pi}\int_\Sigma d^2z\,\,\Tr[\Psi_R{\cal D}_{\bar
z}\Psi_R+\Psi_L{\cal D}_z \Psi_L]
},}
where
$
g\in G,\,\,\,A^a\in{\rm Lie}(H),\,\,\,
\Psi_{L(R)}\in{\rm Lie}(G)-{\rm Lie}(H),\,\,\,{\cal D}_a\equiv \partial_a+
[A_a,\,\,\,].
$
The left and right moving `coset fermions', $\Psi_L$ and $\Psi_R$, are
minimally coupled to the gauge fields such that under:
\eqn\under{\eqalign{g\to&hgh^{-1},\,\,\,
\Psi_{L(R)}\to h\Psi_{L(R)}h^{-1},\cr
A_a\to&h\partial_a h^{-1}+hA_ah^{-1},\,\,\,
{\rm where}\,\,\,h(z,{\bar z})\in H,}}
the model is gauge invariant. It is gauge invariant precisely because $I(g,A)$
is a gauge invariant extension of $I_{WZW}(g)$, while the one--loop
gauge anomalies that arise from the fermions cancel against each other, as the
coupling of the left and right fermions to the gauge fields are identical in
magnitude. The opposite chirality of the fermions then results in a relative
minus sign.

This action has an $N=1$ supersymmetry\ed\shnitzer:
\eqn\Neqone{\eqalign{
&\delta g=i\epsilon_Rg\Psi_R+i\epsilon_L\Psi_Lg\cr
&\delta \Psi_R=\epsilon_R(1-\Pi_0)\cdot(g^{-1}{\cal D}_zg-i\Psi_R\Psi_R)\cr
&\delta \Psi_L=\epsilon_L(1-\Pi_0)\cdot({\cal D}_{\bar
z}gg^{-1}+i\Psi_L\Psi_L)\cr
}}
where $\Pi_0$ is the orthogonal projection of Lie$(G)$ onto Lie$(H)$.

\subsec{$(2,2)$ Supersymmetry from $(1,1)$ Cosets}
The model \SGWZW\ is also to be the definition for the $N=2$ Kazama--Suzuki
models\foot{This action was first
studied in this context  in refs.\ed\
and \ref\Nakatsu{S. Nakatsu, Prog. Theor. Phys. {\bf 87} (1992) 795.}.
In ref.\ed\ this action was used (after twisting) to
do explicit calculations in certain topological field theories.
In refs.\mansi
\ref\mansii{M. Henningson, Nucl. Phys. {\bf B423} (1994) 631, hep-th/9402122.}\
it was used  to study
 important properties of the Kazama--Suzuki models which are more
easily accessible via field theoretic methods. This includes  an
investigation of mirror symmetry for the Kazama--Suzuki models (and their
orbifolds) and a
calculation of the elliptic genus for the $N=2$ minimal models.} for
the reason that
just as in the algebraic construction of Kazama and Suzuki\ref\ks{Y. Kazama and
H. Suzuki, Nucl. Phys. {\bf B321} (1989) 232\semi
Y. Kazama and H. Suzuki, Phys. Lett. {\bf B216} (1989) 112.}, an $N=2$
supersymmetry arises from the $N=1$ above \Neqone\
when the space $G/H$ is K\"ahler\ed:
Taking $T={\rm Lie}(G/H)$ as the complexification of the orthogonal complement
of Lie$(H)$ within Lie$(G)$, the K\"ahler condition translates into the
decomposition $T=T_+\oplus T_- $
where $T_+, T_-$ are complex conjugate representations of $H$,
with
$[T_+,T_+]\subset T_+$, $[T_-,T_-]\subset T_-$, and
$\Tr(ab)=0$ for
$a,b\in T_+$ or $T_-$.
The three pieces of the condition define first an almost complex structure,
then  integrability of this structure,  and finally that the metric is a
$(1,1)$ quadratic form
which is K\"ahler. Continuing to follow ref.\ed, we define $\Pi_\pm$ as the
orthogonal projections of $T$ onto $T_\pm$. The right (left)--moving
`coset fermions'
decompose under this to $\Psi_{R(L),\pm}=\Pi_\pm\Psi_{R(L)}$,
and the fermion action becomes:
\eqn\becomes{
I^F={ik\over2\pi}\int_\Sigma d^2z\,\,\Tr[\Psi_{R,+}{\cal D}_{\bar
z}\Psi_{R,-}+\Psi_{L,+}{\cal D}_z \Psi_{L,-}].
}
Now we can see that there is an R--symmetry (i.e. it does not commute with
the supersymmetry) for each chirality which assigns the charge $\pm1$ to
quantities valued in $T_\pm$ and charge $0$ to $g$ and the gauge fields. The
$N=1$ supersymmetry transformation \Neqone\ can be decomposed into terms of
$\Delta R=\pm1$, which will be our two supersymmetries giving $N=2$, with
parameters
$\epsilon_{R(L),\pm}$:
\eqn\Neqtwo{\eqalign{
&\delta g=i
\epsilon_{R,+}g\Psi_{R,-}+i\epsilon_{R,-}g\Psi_{R,+}\cr
&\delta \Psi_{R,+}=
\epsilon_{R,+}\Pi_+\cdot\left[{\cal D}_{ z}gg^{-1}-i(\Psi_{R,+}\Psi_{R,-}+
\Psi_{R,-}\Psi_{R,+})\right]-i\epsilon_{R,-}\Psi_{R,+}\Psi_{R,+}\cr
&\delta \Psi_{R,-}=
\epsilon_{R,-}\Pi_-\cdot\left[{\cal D}_{ z}gg^{-1}-i(\Psi_{R,+}\Psi_{R,-}+
\Psi_{R,-}\Psi_{R,+})\right]-i\epsilon_{R,+}\Psi_{R,-}\Psi_{R,-}\cr
&\delta({\rm everything}\,\,{\rm else})=0,
}}
and a similar set of $N=2$ transformations on the left--moving side.

\subsec{Lagrangians for $(0,2)$ models}
Recently, in ref.\paperone\papertwo\ it was noted that there are many more ways
of combining the above ingredients to get gauge invariant models, and hence a
larger class of conformal field theories. In particular, for the study of
$(0,2)$ conformal field theories it is possible to preserve the right moving
structure of the Lagrangian, the couplings of the right moving fermions  and
the right action of the gauge group, and hence preserve the right
supersymmetry. On the left hand side,  fermions may be included  with a
priori arbitrary couplings, thus disallowing generally the possibility of a
left supersymmetry. There is now the potential problem that the chiral gauge
anomalies from the left and right do not cancel. This  problem is circumvented
by allowing the possibility to gauge as arbitrary a left action of the gauge
group on $g$ as allowed by group theory. In general, an extension for
$I_{WZW}(g)$ based on this  resulting non--diagonal gauging of the WZW  can be
chosen so as to produce (classical) chiral anomaly terms.  Simply requiring
that the total anomaly from the three sectors vanishes restores gauge
invariance.

This way of producing a $(0,2)$ model by modifying the possible gaugings and
the left fermion coupling allows great freedom in the type of left moving
structures present in the models, as is evident in the prototype example of
this type of construction, the `monopole' theory of ref.\gps, which was shown
to be a heterotic coset in ref.\paperone. In that model,  there is an
additional $SU(2)$ current algebra on the left, which is the
world--sheet manifestation of spacetime rotational invariance. (This model was
presented as the angular sector of a 4D spacetime extremal black hole with
magnetic charge.) Similarly, such symmetries may be found in the heterotic
coset
realization of other spacetime backgrounds\paperone\paperthree.

In general, gauging the following  symmetry of the WZW model
$g\to h^{\phantom R}_Lgh_R^{-1}$
for $(h^{\phantom R}_L,h^{\phantom R}_R)\in
(H^{\phantom R}_L,H^{\phantom R}_R)\subset
(G^{\phantom R}_L,G^{\phantom R}_R)$
is anomalous. This
simply means that one cannot write down an extension of the
WZW model which promotes this symmetry to a local invariance: There will
always be terms which spoil gauge invariance. (This is because of the
Wess--Zumino term; the `metric' term may be simply minimally coupled.)

Knowing that we will get an anomaly, let us choose to write {\sl some}
gauge extension such that under gauge transformations the `anomalous'
piece does not depend upon the group element
$g$. This results in the anomalous piece taking
the form of the standard 2D chiral anomaly. The  {\sl unique}\ref\edagain{E.
Witten, Commun. Math. Phys. {\bf 144} (1992) 191.}\ action is:
\eqn\extendi{\eqalign{I&^{G_k}_{GWZW}(g,A)=\cr
&-{k\over4\pi}\int_{\Sigma} d^2z\,\, \Tr[g^{-1}\d g\cdot g^{-1}\db
g]-{ik\over12\pi}\int_B d^3\sigma\,\,
\epsilon^{ijk}\Tr[g^{-1}\partial_ig\cdot g^{-1}\partial_jg\cdot g^{-1}
\partial_kg]\cr
&+{k\over2\pi}\int_\Sigma d^2z\,\,\Tr[\Azb^R\DG-\Az^L\DBG+\Azb^R g^{-1}\Az^L
g-{1\over2}\{\Az^L\Azb^L+\Az^R\Azb^R\}],}}
where ${A^{R(L)}=A^a t_{a,R(L)}}$.
Under the infinitesimal variation
\eqn\variat{\eqalign{g\to&g+ \epsilon^Lg-g\epsilon^R\cr
\Az^{R(L)}\to& \Az^{R(L)}-\d\epsilon^{R(L)}-[\Az^{R(L)},\epsilon^{R(L)}]\cr
\Azb^{R(L)}\to& \Azb^{R(L)}-\db\epsilon^{R(L)}
-[\Azb^{R(L)},\epsilon^{R(L)}],\cr
{\rm for}\,\,\, \epsilon^{R(L)}=&\epsilon^a t_{a,R(L)}}}
the variation is
\eqn\vary{
\eqalign{
\delta I(g,A)&={k\over4\pi}\Tr[t_{a,R}\cdot t_{b,R}-t_{a,L}\cdot
t_{b,L}]\int_\Sigma d^2z \epsilon^{(a)}F^{(b)}_{z\bar z}\cr
{\rm where}\,\,\,t_{a,L(R)}\in &{\rm Lie}(H),\,\,\, {\rm and}\,\,\,
F^{(b)}_{z\bar z}\equiv \d\Azb^{(b)}-\db\Az^{(b)}.
}}
Notice in particular that for the popular diagonal gaugings of WZW models
this variation is zero and the action reduces to the familiar one.

Turning to the right moving Majorana--Weyl  fermions,
 it is sufficient to minimally couple them as  coset fermions
to the gauge fields:
\eqn\IFR{\eqalign{I_F^R(\Psi_R,A)=&{i k\over4\pi}\int_\Sigma\,\,
\Tr[\Psi_R{\cal D}_{\bar z}\Psi_R]\cr
{\rm where}\,\,\,{\cal D}_{\bar z}\Psi_R&=\db\Psi_R+
[\Azb^R,\Psi_R],\,\,\Psi_R\in{\rm Lie}(G)-{\rm Lie}(H).}}
This model is classically gauge invariant under:
\eqn\class{\eqalign{\Psi_R\to h^{\phantom R}_R \Psi_Rh_R^{-1}\,\,\,
{\rm and}\,\,\,
A^R\to h^{\phantom R}_Rd h_R^{-1}+h^{\phantom R}_RA^R h_R^{-1} }}

There are $D=$ dim$(G)-$dim$(H)$ fermions $\psi^i_R$ in $\Psi_R$, all coupled
with charges derived from the generators $t_{a,R}$.
The chiral anomalies appear at one loop and  are\foot{Here and for the
remainder of this paper, it is implicit that a consistent regularisation scheme
has been chosen for calculation of the fermion anomalies, and such that the
normalisation of the anomalies is chosen to be of this simple
form.}:
\eqn\Ranomalies{{1\over4\pi}\Tr_{\rm Ad}[t_{a,R}\cdot t_{b,R}]\int_\Sigma d^2z
\epsilon^{(a)}F^{(b)}_{z\bar z}.}
(Note here  the absence of $ k$, which plays the role of $1/\hbar$. This
 really is a one loop effect.) Here $\Tr_{\rm Ad}$ means the trace in the
adjoint representation.

It is a natural choice to add
$D=$dim$(G)-$dim$(H)$ left moving Majorana--Weyl fermions
 with arbitrary values of the minimal couplings\foot{It is important to realize
that this choice is not necessary. Indeed, the most natural way to proceed
beyond this would be to add more left moving fermions, in order to increase the
total central charge of the left--moving sector. This would then immediately
give rise to the possibility of smaller  spacetime gauge groups for
model building. For now, we shall merely note this in passing but hope to
return to
this important point in the not too distant future.}. To
be precise, arrange them into a fundamental vector
$\Lambda_L=\{\lambda^i_L\}$ of the
group 
$SO(D)_L$ which acts on them as a global symmetry, and minimally couple
them to the
$H_L$ subgroup with generators $Q_{a,L}$ in this fundamental representation:
\eqn\IFL{I_F^L(\lambda^i_L,A)={ik\over4\pi}\int_\Sigma d^2z \,\,
\Lambda^T_L(\d+\sum_a\Az^aQ_{a,L})\Lambda_L.}

Their chiral anomalies
appear at one loop and are:
\eqn\Lanomalies{-{1\over4\pi}{\widetilde\Tr}[Q_{a,L}\cdot Q_{b,L}]\int_\Sigma
d^2z \epsilon^{(a)}F^{(b)}_{z\bar z}.}
(Note again the absence of $k$. Also note the minus sign relative to
\Ranomalies, due to the opposite chirality. Here $\tilde\Tr$ is the
trace in the fundamental representation of $SO(D)$.)

So if we add together the three actions \extendi, \IFR\ and \IFL, we get
a gauge invariant  model if we ensure
that all of the  anomalies (classical and
quantum) cancel: \eqn\cancelone{k\Tr[t_{a,R}\cdot
t_{b,R}-t_{a,L}\cdot t_{b,L}]+\Tr_{\rm Ad}[t_{a,R}\cdot
t_{b,R}]-{\widetilde\Tr}[Q_{a,L}\cdot Q_{b,L}] =0.}

Note that $(0,2)$ supersymmetry is  still present given that the parent
$(N=1)_R$ is preserved (see equation \Neqtwo), along with the structure of
$G/H$ as seen by the right movers.

\subsec{Bosonization}
So far, the   model as written  is gauge invariant when we take into account
the
one--loop effects. To write a {\sl classically}
gauge invariant action it is necessary to {\sl bosonize} the fermions. The
bosonic action equivalent to $I_R^F+I_L^F$ is {\sl classically} anomalous, and
is
easily seen to be\papertwo\ an  $SO(D={\rm dim}G-{\rm dim}H)$ WZW (at level 1)
gauged anomalously with different embeddings of
$H$ in $SO(D)$ on the left and on the right:
\def\tg{{\tilde g}}
\def\th{{\tilde h}}
\eqn\choices{
\eqalign{{\tilde g}\to {\tilde h}^{\phantom R}_L{\tilde g}{\tilde h}_R^{-1}
\,\,\,{\rm for}\,\,\,{\tilde g}\in SO(D)\,\,\,{\rm and}\,\,\,
(\th_L,\th_R)\in(H_L,H_R)\subset(SO(D)_L,SO(D)_R)
}
}
The $(H_L,H_R)$ are generated by $(Q_{a,L},Q_{a,R})$. Choose the
$Q_{a,R}$ such that when acting on the $\psi_R^i$'s in the fundamental
representation of $SO(D)$ they are equivalent to the $t_{a,R}$ acting on
the $\psi^i_R$ in the coset fermion $\Psi_R\in{\rm Lie}(G)-{\rm Lie}(H)$.
This will ensure that the right moving fermions are correctly coupled and
preserve the (now hidden) $N=2$ on the right.

Then the bosonic action equivalent to the interacting fermions is just an
action of the form \extendi\ (with level 1), which yields the classical
anomalies:
\eqn\anon{
{1\over 4\pi}{\widetilde\Tr}
[Q_{a,R}\cdot Q_{b,R}-Q_{a,L}\cdot Q_{b,L}]
\int_\Sigma d^2z \,\,\epsilon^{(b)}F_{z\bar z}^{(a)}.
}
So canceling this against the anomaly of the $G/H$ bosonic model
(and recalling from the above paragraph that
${\tilde\Tr}[Q_{a,R}\cdot Q_{b,R}]=\Tr_{\rm Ad}[t_{a,R}\cdot t_{b,R}]$), we
recover \cancelone\ as the condition for a consistent model.

So finally we can write a classically gauge invariant analogue of
\SGWZW\ which realizes a $(0,2)$ conformal field theory  written as the sum of
two gauged \WZW\ models which are
separately anomalous:
\def\DTG{{\tg^{-1}\d \tg}}\def\DBTG{{\db \tg\tg^{-1}}}
\def\TTr{\widetilde\Tr}
\eqn\final{\eqalign{&I^{(0,2)}=I_{GWZW}^{G_k}(g,A)+
I_{GWZW}^{SO(D)_1}(\tg,A)\cr
&=I_{WZW}^{G_k}(g)+ I_{WZW}^{SO(D)_1}(\tg)+\cr
&+{k\over2\pi}\int_\Sigma d^2z\,\,\Tr[\Azb^R\DG-\Az^L\DBG+\Azb^R g^{-1}\Az^L
g-{1\over2}\{\Az^L\Azb^L+\Az^R\Azb^R\}]\cr
&+{1\over2\pi}\int_\Sigma d^2z\,\,\TTr[\Azb^R\DTG-\Az^L\DBTG+\Azb^R
\tg^{-1}\Az^L
\tg-{1\over2}\{\Az^L\Azb^L+\Az^R\Azb^R\}],
}}
where $ D={\rm dim}(G)-{\rm dim}(H)$ and so
the heterotic coset is realized as:
$\left[G_k\times SO(D)_1\right]/ H$ with the gauged symmetry \variat\ and
\choices\ and subject to \cancelone.

\subsec{The $(0,2)$ minimal models}
For most of the sections in this paper, we will consider in detail the case
$G=SU(2)$ with $H=U(1)$. These models are therefore the analogue of the $(2,2)$
minimal models and will accordingly be referred to as the
$(0,2)$ minimal models. Specializing some of the previous formulae to this
case,   we will use
\eqn\parameters{\eqalign{
h^{\phantom R}_L=\e{i\epsilon\alpha\sigma_3/2},\,\,
h^{\phantom R}_R=\e{-i\epsilon\sigma_3/2},\,\,
\th^{\phantom R}_L=\e{i\epsilon Q\sigma_2},\,\,
\th^{\phantom R}_R=\e{-i\epsilon\sigma_2},\,\,
{\rm and}\,\,\tg=\e{i\Phi\sigma_2},
}}
and we have for the heterotic coset model in bosonic
form:
\eqn\minimal{\eqalign{&I^{(0,2)}_{\alpha,Q}=I_{WZW}(g\in SU(2),k)\cr
&+{k\over2\pi}\int d^2z\Tr\Biggl[\Azb g^{-1}\d g+\alpha\Az\db g
g^{-1}-\alpha\Azb g^{-1}\Az g-{1\over2}(1+\alpha^2)\Az\Azb \Biggr]\cr
&+{1\over4\pi}\int d^2z \left[\d\Phi\db\Phi- 2\Azb\d\Phi
-2Q\Az\db\Phi+(1+Q)^2\Az\Azb     \right],
}}
which is invariant under the following infinitesimal
gauge transformations:
\eqn\mingauge{\eqalign{
g\to g+{i\over2}\epsilon(\alpha \sigma_3g+g\sigma_3),\,\,\,
\Phi\to\Phi+(Q+1)\epsilon,\,\,{\rm and}\,\,
A_a\to A_a+\partial_a\epsilon,}}
subject to the anomaly cancelation condition
\eqn\anom{k(1-\alpha^2)=2(Q^2-1).}

The theory of the $2\pi$ periodic boson $\Phi$ is the bosonised fermions and
is equivalent to:
\eqn\theferm{I_{F}={i k\over4\pi}\int
d^2z\left\{\Tr_{\rm Ad}[\Psi_R{\cal D}_{\bar
z}^R\Psi_R]+\Lambda_L^T(\d+Q\cdot\Az)\Lambda_L \right\}.}
The left--moving fermions are coupled to the $U(1)$ gauge fields with
 charge $Q$, whereas the right--movers are coset fermions with their
charges determined by geometry.
The cases $\alpha=\pm1$, for which (due to \anom)
$Q=\pm1$ correspond to
the familiar diagonal models, which is the family of $(2,2)$ minimal models.
The case $\alpha=0$ is the charge $Q$
`monopole' theory of ref.\gps, and will henceforth
be the focus of this paper.
Other $\alpha$ are restricted to being integer, in order to furnish
a faithful representation of a $U(1)$ subgroup of $SU(2)$. However,
the anomaly equation prohibits such
solutions\foot{Note that the  restrictions on analogous parameters
in the case of heterotic cosets based upon non---compact groups are
not so severe\paperone\papertwo. Also, the use of rational $\alpha$
for compact groups seems to give results consistent with conformal
invariance, at least at one--loop\paperthree.}.

The central charge (on both the left and right) of a $(0,2)$
minimal model of level $k$ is given
by the familiar formula
\eqn\central{c={3k\over k+2}.} This follows from the simple fact that
the gauging of a $U(1)$ subgroup effectively
removes one bosonic degree of freedom, whose central
charge is equal to that of the two fermions added for supersymmetry (on the
right) or of current algebra fermions (on the left). So the central charge of
the
$SU(2)$ contribution to the model is all that remains.

\newsec{Elliptic Genera for the $(0,2)$ minimal models}
Of all quantities which exist for supersymmetric models, those which correspond
to topological invariants of an associated geometry (when there is such a
geometry), or more generally an index of some type, are usually most
accessible.
Amongst such quantities is the elliptic genus\ref\genuselliptic{A. Schellekens
and N. P. Warner, Phys, Lett. {\bf B77}
 (1986) 317; Nucl. Phys. {\bf B287} (1987) 317\semi K. Pilch, A. Schellekens
and N. P. Warner, Nucl. Phys. {\bf B287} (1987) 362.}\ref\edelliptic{E. Witten,
\cmp{109}{1987}{525}; {\sl `The Index of the Dirac Operator in Loop Space'},
in {\sl `Elliptic Curves and Modular Forms in algebraic Topology'}, ed. P.S.
Landweber,Springer--Verlag, 988.}\ref\moreelliptic{O. Alvarez, T. Killingback,
M. Mangano and P. Widney, \cmp{111}{1987}{1}; Nucl. Phys. {\bf B} (Proc.
Suppl.) {\bf 1A} (1987) 189.}.
Given a $(0,2)$ theory with at
least a $U(1)$ current algebra on the left (which commutes with the $N=2$
supersymmetry), the partition function may be written as follows:
\eqn\partit{\eqalign{Z(q,\gamma_L,\gamma_R)&=\Tr_{\cal
H}\left[(-1)^{F_L}q^{H_L}\exp{(i\gamma_LJ_{0,L})}
(-1)^{F_R}q^{H_R}\exp{(i\gamma_RJ_{0,R})}\right]}.}

The elliptic genus arises when we consider the restriction of  the
Ramond sector
partition function to
the case $\gamma_R=0$. Then the right--moving sector contains only the
quantity $\Tr(-1)^{F_R}q^{H_R}$. Due to the presence of supersymmetry,  this
quantity is only ever 1 or 0 when evaluated on sectors of the Hilbert space
$\cal H$.
By not grading states  according to  their $U(1)_R$ charge (i.e. putting
$\gamma_R=0$),
 the boson and fermion pairs
at  each non--zero mass
level contribute equally to the sum, but with a relative minus sign  and thus
contributions from sectors with $H_R\neq0$ will sum to  zero. Only the
right moving   ground
state sectors  (those with $H_R=0$) can  contribute, due to the presence of
unpaired states.

The elliptic genus is thus:
\eqn\ellipic{\eqalign{Z(q,\gamma_L)&=
\Tr_{{\cal H}(H_R=0)}\left[(-1)^{F_L}q^{H_L}\exp{(i\gamma_LJ_{0,L})}\right]}}
which will encode for us all of the information about the states in the
left moving sector which couple to the right moving Ramond ground states.

It was in ref.\edminimal\ that the elliptic genera for the ($A$--series of the)
  $(2,2)$ minimal models was calculated in their Landau--Ginsburg formulation
enabling  comparison
to those of their algebraic
formulation\ref\difran{P. Di Francesco and S. Yankielowicz, Nucl. Phys.
 {\bf B411} (1994)
584, hep-th/9306157\semi P. Di Francesco and S. Yankielowicz, Nucl. Phys. {\bf
B409} (1993)
186, hep-th/9305037.}. This provided further evidence
to support the conjecture that the Landau--Ginsburg models did indeed flow to
the minimal models at their fixed point. In ref.\mansi, the elliptic genera of
the ($A$--series of the)  $(2,2)$ minimal models was calculated using path
integral techniques based upon the $(2,2)$ supersymmetric gauged WZW models
discussed in the previous section. It is this approach which we will follow
most closely in order to calculate the elliptic genera of our $(0,2)$ minimal
models. The machinery will need little modification to be applied here, as
indeed our models are also based upon supersymmetric gauged WZW models, and
furthermore the right--moving sector is identical to those of the $(2,2)$
models.

\subsec{Global $U(1)_L$ symmetry}
To begin with we need to identify the appropriate left--moving $U(1)$ symmetry
whose global component we will use to grade all fields in the model, in
preparation for the twisting procedure.
In the $(2,2)$ case, this was the $U(1)$ of the left $N=2$, which commuted with
the right $N=2$. For this case we of course still need the latter constraint.
This $U(1)$ will act upon the fermions of our model (which we have arranged
into complex fermions $\lambda_L$ and $\psi_L$), and the bosonic fields $g$.

Following ref.\mansi\ we postulate the following changes under global $U(1)_L$:
\eqn\changesone{\eqalign{\delta\lambda_L=i\epsilon c_L\lambda_L;\,\,
\delta\psi_R=i\epsilon c_R\psi_R;\,\,
\delta g={i\over2}\epsilon(x_L\sigma_3g+x_Rg\sigma_3)
}}
and similarly for the complex conjugate fields
$\bar\psi_R$ and $\bar\lambda_L$. The quantities $c_L,c_R,x_L$ and $x_R$ are
charges to be determined.
The requirement that  this transformation commutes with the right supersymmetry
fixes $-x_R=c_R$. Consistency with a left  supersymmetry can no longer be a
requirement here as in the $(2,2)$ case, because it is not present in a $(0,2)$
 model. So we do
{\sl not} have the condition $x_L=c_L-1$,
which arose in ref.\mansi. However it
transpires that  $c_L$ turns out to be free and can be chosen to set the
overall normalisation of the charges of $U(1)_L$.

Now we wish for our complete action to be invariant under this global $U(1)_L$
symmetry. Structurally the situation is almost identical to the 2D gauge theory
 considerations we made earlier, when we constructed the coset:
the bosonic sector ($I_{WZW}(g)$ and its gauge extension) is not invariant
under \changesone, and under those transformations, produces the `classical
anomaly' $\epsilon kx_R$. Meanwhile, the fermions \theferm\ are
classically invariant under \changesone, but produce one--loop anomalies
$2\epsilon(c_R-c_LQ)$. The anomaly cancelation equation is then
$2(c_LQ-c_R)=kx_R$.
We have  four parameters ($x_L,x_R,c_L,c_R$) and two equations. It is
convenient to  use a gauge transformation to set $x_L=-x_R=x$, and putting
$c_R=-x_R$ (determined  above)
gives $x=-{2c_LQ/(k+2)}$.
So $c_L$ is a free parameter here and  is just a normalisation of the  $U(1)_L$
charges, which we are  always free to fix if there are no other constraints. If
we rescale such that $c_L={1/(2Q)}$  then:
\eqn\changestwo{\eqalign{\delta\lambda_L= {i\epsilon \over
2Q}\lambda_L,\,\,\,
\delta\psi_R={i\epsilon\over(k+2)}\psi_R,\,\,\,{\rm and}\,\,\,
\delta g=-{i\epsilon\over2(k+2)}(\sigma_3g+g\sigma_3),
}}
and similarly for the complex conjugates $\bar\psi_R$ and $\bar\lambda_L$.
This represents a scale choice which produces the  normalisation for
the charges of the right--moving sector of the  minimal models as set out in
refs.\edminimal\mansi.

\subsec{An Index and Deformation to Free Theory}
The elliptic genus, equation \partit\ with $\gamma_R=0$, may be regarded as a
path integral evaluated  on the torus with twisted boundary conditions.
To be more precise, let us define our torus as usual starting with the lattice
obtained by folding the $x^1-x^2$ plane according to $x^1\to x^1+m$, $x^2\to
x^2+n$ for $m,n\in \IZ$. The torus with modular parameter $\tau$ is supplied
with complex coordinate via $z=x^1+\tau x^2$. We will chose $x^2$ as the `time'
direction and $x^1$ as `space'.

The elliptic genus is then defined as the path integral
\eqn\path{\eqalign{Z_{Q}^{(0,2)}(\tau,\gamma)=
\int_{\rm torus}\!\!\!{\cal D}g{\cal D}\Psi_R
{\cal D}\Lambda_L{\cal D}\Az{\cal D}\Azb\,
\e{-I^{(0,2)}_{Q}(g,\Psi_R,\Lambda_L)}
}}
with periodic boundary conditions in the space direction and for the time
direction there is the $U(1)_L$ twist:
\eqn\twistL{\eqalign{
g(x^1,x^2+1)=& \e{-{i\gamma\sigma_3\over2(k+2)}}g(x^1,x^2)
\e{-{i\gamma\sigma_3\over2(k+2)}}\cr
\lambda_L(x^1,x^2+1)=& \e{i\gamma
\over 2Q}\lambda_L(x^1,x^2)\cr
\psi_R(x^1,x^2+1)=& \e{i\gamma\over(k+2)}\psi_R(x^1,x^2)\cr
A_a(x^1,x^2+1)=& A_a(x^1,x^2)\cr
}}

Now as the elliptic genus is a supersymmetry  index, there should exist smooth
deformations of the system which preserve the supersymmetry and hence keep this
quantity unchanged. Indeed it was this philosophy which was adopted in
ref.\edminimal\ in order to calculate the  elliptic genus for the
Landau--Ginsburg models. By safely deforming the theory to the weak coupling
regime, a successful computation could be carried out.

In ref.\mansi,
this procedure was carried out for  the gauged WZW formulation of
the minimal models. The deformation appropriate to the problem was identified
and the calculation reduced to a weak coupling problem. Once the $U(1)_L$
charges of the constituent fields were identified, the problem was reduced to
the free field computation of ref.\edminimal.
Here, the same procedures follow. The structure of the right--moving sector is
identical to that of ref.\mansi\ and therefore the operator with which to
deform the theory safely to weak coupling is the same. As we have identified
the global $U(1)_L$ charges in the previous section, there only remains the
task of performing again the free--field computation of ref.\edminimal, taking
into account the contributions from the fermionic and bosonic modes in a
Hilbert space approach.

This results in the
following expression for the
elliptic genera of the $(0,2)$ minimal models.
\eqn\ellipticgenus{\eqalign{
Z^{(0,2)}_{Q}(q,\gamma,0)=
\e{-i\gamma {k(Q-1)\over2(k+2)}}\cdot
{
1-\e{i{\gamma\over2Q}}\over
1-\e{i{\gamma\over k+2}}
}
\prod_{n=1}^\infty
{
(1-q^n\e{i{\gamma\over2Q}})
(1-q^n\e{-i{\gamma\over2Q}})\over
(1-q^n\e{i{\gamma\over k+2}})
(1-q^n\e{-i{\gamma\over k+2}})
}.
}}

\subsec{Properties of the Elliptic Genera}
In this section we note some important properties of the elliptic genera
which we computed above. This serves as a useful warmup for the
case of the complete partition functions, computed later in this paper. Indeed,
some of the information extracted here will prove to be useful in those later
computations.

First note that by expanding the expression \ellipticgenus\ one can guess the
following formula:
\eqn\oohlookone{
Z^{(0,2)}_{Q}(q,\gamma,0)~=~\sum_{r=0}^{Q-1} (-1)^r \X_{Q-1+2Qr;k}
}
where $\X_{l;k}$ is the character for the highest weight unitary
irreducible representation with isospin $l/2$ of the level $k$ $SU(2)$ affine
Lie algebra.
This relation
 is proven in Appendix A, using a convenient representation of the
elliptic genera in terms of theta functions.

This explicitly shows the emergence of the `physical' $SU(2)$ affine Lie
algebra
on the left since the gauging  leaves the $SU(2)_L$ symmetry untouched:
$g\to g\exp({i\epsilon{\sigma_3/2}})$. We have for the heterotic coset at level
$k$ a family of $Q$ level $k$ affine $SU(2)$ characters on the left.
Recall that the information content of the elliptic genus is precisely about
all of the states from the left that couple the  (supersymmetric) right moving
sector's Ramond ground states. In this case, the affine $SU(2)$ encodes
 the spacetime rotation invariance of the `monopole' (of charge $Q$) theory of
\gps. There, the left moving fermions carry the spacetime $U(1)$ monopole
field of the heterotic string magnetic black hole background.

Turning to the formula \ellipticgenus\ for the elliptic genera again,
we see that in order to have a finite number of terms in the $q^0$
level of the elliptic genus expansion, one must have that $Q$ is an integer, as
one would expect from the $U(1)$ monopole interpretation of ref.\gps.
This is of course a quite physical requirement from the CFT point of view,
  saying
 that we have a finite number, $Q$, of highest
weight vacuum states.

\def\tg{{\tilde g}}

\newsec{The Partition Functions of the $(0,2)$ Minimal Models}
In the previous section we calculated the elliptic genera for the $(0,2)$
minimal models. In many formulations of a superconformal field theory, the
calculation of this quantity is the closest one can get to the full partition
function of the theory. In this case, it turns out that we can go much further,
and calculate the full partition function for  these models.

Our model is a manifestly left--right asymmetric combination of two gauged WZW
models, and it is not obvious just how to make sense of the task of
constructing
its spectrum, given these unusual couplings, although
a number of statements may be made given that there is a familiar $SU(2)$
current algebra  on the left, as  we saw in the
last section. Even with this knowledge, it is not an easy task to construct
the modular invariant combination of characters which gives the partition
function of these
models when the asymmetry is present.
It is therefore of great comfort to note that progress can be made by using our
field theory intuition to study the path integral, and answer difficult
algebraic questions.

\subsec{A Change of Variables}
Motivated by the gauge transformations under which the model \minimal\
is invariant,
 a little thought suggests that the
following changes of variables
\eqn\change{
\eqalign{\Az&\to\sqrt{2}\d\phi_L,\,\,\,\Azb\to\sqrt{2}\db\phi_R;\cr
g&\to g\e{-i{\sigma_3\over2\sqrt2}\phi_R},\,\,\,
\Phi\to \Phi+\sqrt{2}(Q\phi_L+\phi_R)}} might be interesting,
for the simple reason that
they would formally uncouple the action of the gauge symmetry from the original
variables, and
put them entirely on the $\phi_{L(R)}$ fields:
$\delta\phi_{L(R)}=\epsilon/\sqrt2$.

The change of variables for all the fields except the gauge fields are
harmless\foot{Notice that  this simple change of variables for the gauge fields
 fixes us to only considering gauge configurations which can be deformed to the
identity, i.e. the trivial holonomy sector. Therefore we are really only
working on a world sheet with the topology of the sphere. Later, we have to
take other topologies and gauge configurations into account.}.
However, those for the gauge fields require a non--trivial Jacobian to be
computed\foot{Note that for subgroup $U(1)$ the usual corrections proportional
to the quadratic Casimir $C_H$  due to the change of variables on the fermions
and from the Jacobian of the covariant derivative\ref\poly{A. M. Polyakov and
P. B. Wiegman, \plb{131}{1983}{121}; \plb{141}{1984}{223}.}, do not appear.},
which we shall replace with an action for anticommuting ghosts in the
standard way:
\eqn\standard{\eqalign{
\D\Az\D\Azb&=\D\phi_L\D\phi_R\det[\d]\det[\db
]=\cr
&=\D\phi_L\D\phi_R\int\D b\D c\D{\bar b}\D{\bar c}\,\exp \left[i\int d^2z\,(
b\db c+{\bar b}\d{\bar c})\right] }}

After some algebra, and much strategic use of the anomaly equation to simplify
expressions
we find that the change of variables does indeed formally decouple the systems
from one another:
\eqn\decoupled{I^{(0,2)}_{\alpha,Q}=I_{WZW}^{SU(2)_k}+I_{WZW}^{SO(2)_1}+
I_{WZW}^{SO(2)_{-(k+2)}}+\int d^2z\,(b\db c+{\bar b}\d{\bar c}).}
We have written the final result somewhat suggestively. Let us examine the
terms.
The first is just the pure $SU(2)$ WZW, while the second term is simply the
kinetic term for the free periodic boson (the bosonised fermions),
${1\over4\pi}\int\d\Phi\db\Phi$.
The third term is actually the theory:
\eqn\props{
I=-{(k+2)\over4\pi}\int d^2z (\d\phi_L-\d\phi_R)(\db\phi_L-\db\phi_R)).
}
The idea is that we will fix a gauge by setting $\d\phi_L=0$ (i.e. $\Az=0$).
This then gives us a  level $-(k+2)$ $SO(2)$ WZW theory in the variable
$\phi_R\equiv\phi$. (Notice that this level is somewhat arbitrary, except for
the sign,
 but for internal
consistency if  the $\Phi$ theory  has level 1 then this is the right
interpretation  for this theory.)
 This arbitrariness  is lifted for non--abelian $H$.
Notice the analogy with the diagonal case. Similar changes of variables have
been found in that case which allows such a
decoupling\cosetgWZW\ref\bastianelli{F. Bastianelli,
\npb{361}{1991}{555}.}.

Note here that the
 central charge of the models can be computed giving the same  result
 as before,\central, with
slightly differing origins of the cancelations of the contributions which
are not from $SU(2)_k$. This time, each $SO(2)$ theory contributes 1 to the
central charge, while the ghosts contribute $-2$.

Of course, this decoupling into a sum of WZW's does not complete the story. The
gauge symmetry must still be present somehow, imposing conditions upon the
complete theory in order to recover the coset.
Such  constraints will arise from the BRST symmetry of the system, which may be
derived using the  methods in ref.\bastianelli.
Using the following normalisation for WZW affine currents:
\eqn\covs{J_a=k\Tr[t_ag^{-1}\d g],\,\,\, {\bar J}_a=k\Tr[t_a\db gg^{-1}]}
for a generic model $g$ at level $k$, (for $SU(2)$ we use
$t_3=i{\sigma_3/2}$), and  denoting  the left (right) currents for the
$\Phi$ and $\phi$ theories as
$J$ ($\bar J$) and $I$ ($\bar I$) respectively, we derived the BRST currents:
\eqn\currents{\eqalign{{\bar J}_{BRST}={{\bar c}\over 2\pi}
\left({\bar J}_3+{\bar J}+{\bar I}\right),\,\,\,
{ J}_{BRST}={ c\over2\pi}\left(Q{ J}-{I}\right).}}
These BRST currents give rise to the nilpotent BRST charge operators:
\eqn\nil{
\eqalign {{\bar Q}_{BRST}=&\oint{dz\over2\pi
i}: {\bar c}\left({\bar J}_3+{\bar J}+{\bar I}\right):\cr
{Q}_{BRST}=&-\oint{dz\over2\pi i}:
{ c}\left(Q{ J}-{ I}\right):
}
}
which allow us to define physical states via the cohomological
problem
\eqn\brstcoho{\eqalign{Q_{BRST}|{\rm Phys}\!>=0;\,\,\,\,\,
|{\rm Phys}\!>\sim|{\rm Phys}\!>+Q_{BRST}|{\rm Anything}\!>}.}
Here $|{\rm Phys}\!>$ is made up of a direct product of states
from each of the
decoupled sectors in \decoupled.
That this procedure gives rise to the same physical content as the coset
construction is a problem which needs careful consideration. The question of
which representations of the constituent sectors which may appear must be
answered. These problems were studied for the diagonal case in the language of
\brstcoho\ in refs.\cosetgWZW\ and \hwangco. The results may be simply stated
as follows\hwangco: For $J^a$ in $H$ and $J^i$ in the Cartan sub--algebra
of $H$, the (relative) cohomological problem (i.e. for states annihilated by
$b^i_0$) reduces to the physical state
condition $J^{a}_0|{\rm Phys}\!>=0$ (on the right or left), which is
equivalent to the coset construction if integrable representations of the
ungauged WZW sector appear ($SU(2)_k\times SO(2)_1$ in our case here) and if
the non--unitary sector coming from the gauge fields
($SO(2)_{-(k+2)}$ in our case) contains no null vectors.
We shall implement these conditions later in this paper.

\def\JJ{{\rm J}}
\def\JJb{{\bar {\rm J}}}
\def\thetab{{\bar \theta}}
\def\P{{\rm P}}
\def\Pb{{\bar {\rm P}}}
\def\qb{{\bar q}}
\def\Lb{{\bar L}}
\subsec{From the Sphere to the Torus: `Continuous Orbifolds'}
In the previous subsection we derived the  powerful result that the
non--diagonal action defining our models may be decoupled into a set of
free WZW models (diagonal) together with a left--right asymmetric BRST
system. The problem of identifying physical states  then became one of
projecting onto the gauge invariant subspace of the starting Hilbert space,
which is just a product of states from WZW models. This  is great progress, but
it does not yet tell us anything about the consistent coupling between left and
right sectors of the theory, as we have not yet put the theory on the torus.

The partition function of the sum of theories \decoupled\ may be written as
(ignoring any $U(1)$ symmetries which commute with the Hamiltonian)
\eqn\bigZ{Z=\Tr_{{\cal H}_0} q^{L_0-{c\over24}}{\bar q}^{{\bar L}_0-
{\bar c\over24.}}}
The big Hilbert space ${\cal H}_0$ is of the big theory
$I_{WZW}(g)+I_{WZW}(\tg)+I_{WZW}(\phi)+I_{\rm ghosts}$ and $L_0,{\bar L}_0, c$
and $\bar c$ refer to the appropriate quantities when acting in each subsector
of the big Hilbert space ${\cal H}_0$. We obtained physical state information
on the
sphere in the last section by projecting ${\cal H}_0$ onto the subspace of
states in the gauge--BRST cohomology.
To begin to implement this at the level of partition functions we  need to
define a projection operator to insert into the $\Tr$ above. In a standard
orbifold--like
 procedure, with a symmetry under a finite discrete group $H=\{h_i\}$
to take into account, the projection operator would be
\eqn\projects{P={1\over|H|}\sum_ih_i.}
Clearly in our case, the appropriate operators would
be\foot{Note that these two
different projections would be tantamount to twisting in orthogonal directions
on the world sheet. See later in the text.}
\eqn\projection{\eqalign{\P={1\over2\pi}\int\! d\theta\,\e{i\theta\JJ_0},\,\,\,
\Pb={1\over2\pi}\int\! d\thetab\,\e{i\thetab\JJb_0},}}
where
$\JJ_0=QJ_0-I_0,\,\,{\rm and}\,\,
\JJb_0={\bar J}^3_0+{\bar J}_0+{\bar I}_0$.
However as we are studying string theory, we know that we must be much more
 careful.
Considering the theory on the torus, the possibility of `twisted sectors'
should be taken into account.
\def\f#1#2{{[g,\Phi,\phi](#1,#2)}}
On the torus, we have the standard boundary conditions:
\eqn\torusone{\eqalign{\f{x^1+1}{x^2}=\f{x^1}{x^2}\cr
\f{x^1}{x^2+1}=\f{x^1}{x^2}.}}
(We use the notational triplet $\f{x^1}{x^2}$.)
The twisted sectors are precisely the fields with the boundary conditions:
\def\g#1{\left[g\e{#1 i{\sigma_3\over2}},
\Phi+(Q+1)#1,
\phi+#1\right](x^1,x^2)}
\eqn\torustwo{\eqalign{\f{x^1+1}{x^2}=\g{\theta_1}\cr
\f{x^1}{x^2+1}=\g{\theta_2},}}
where $\theta_{1,2}$ are arbitrary elements of the gauge algebra.
The twisted sector partition function is then:
\eqn\twisted{\eqalign{Z(\tau,\theta,\thetab)=\Biggl\{
\Tr_{{\cal H}_{SU(2)_k}}&\,\,
q^{L_0}\,\,\e{-2\pi i
\thetab {\bar J}^3_0}\qb^{\Lb_0}\cdot\,\cr
\Tr_{{\cal H}_{S0(2)_1}}&\,\,\e{2\pi iQ \theta J_0}
q^{L_0}\,\,\e{-2\pi i \thetab
{\bar J}_0}\qb^{\Lb_0}\cdot\,\cr
\Tr_{{\cal H}_{S0(2)_{-{(k+2)}}}}&\,\,
\e{-2\pi i \theta I_0} q^{L_0}\,\,\e{-2\pi
 i \thetab {\bar I}_0}\qb^{\Lb_0}\Biggr\}}}
These type of twisted partition functions  are going to be mapped into one
another under modular transformations in the usual way.
In the standard orbifold language we would construct the modular invariant
partition function by summing over all of the twisted sectors. Here we
integrate
instead. By integrating over sectors twisted over both cycles, we see that we
include both types of  projections \projection\ naturally.
Our result after doing the orbifold is then:
\eqn\partend{Z_{ Q}^{(0,2)}
=\int\!d\theta\, d\thetab\,Z(\tau,\theta ,\thetab )}

Some comments pertaining to the relation to the field theory picture of last
section are in order here.  The twisted sectors of the partition function
language here and the non--trivial holonomy sectors of the gauge fields which
were (knowingly) ignored in the last section are of course related. The twisted
sectors arise when it is  taken into account that a field  going around a
cycle of the torus can return  to a field which is related to the original
field we thought of {\sl up to an arbitrary gauge transformation.} When
on the sphere in the last section we gauge fixed by fixing
 $\phi_L$ to be  a constant,
 but  by parametrising the gauge fields in the way done in eqn.\change,
we took into account only gauge configurations which may be connect locally,
i.e. within each twisted sector. Therefore
we only gauge fixed such transformations also. Nothing was done in the previous
section about the
possibility of jumping between gauges which are distinct because of the
non--trivial topology of the torus. Gauge transformations characterized by
\torustwo, were still unaccounted for. This is reminiscent of taking into
account the holonomy of the gauge field\gk. In this case of the minimal models,
where  we have that the symmetry group is abelian, the boundary conditions
\torustwo\  are a consistent set of boundary conditions.

More generally, when the subgroup is not abelian, things become more
complicated. We have to restrict to the maximal torus of the subgroup which in
the field theory language we can always do because the holonomy defines a map
from the fundamental group of the torus to
 the subgroup, which can always be
conjugated into the Cartan sub--algebra.
Some care must be exercised to make
sure that the conjugation and Weyl freedoms left over can be properly accounted
for, if working directly with the holonomy, as in ref.\gk.
In order to define the continuous orbifold (defined here) correctly,
we note\hwangco\ that the appropriate projection operators
generalising \projection\ contain only
$J_0^i$, i.e. those currents in the Cartan sub--algebra of $H$, due to the
requirement of  compatibility with the BRST conditions \brstcoho. Therefore,
the care needed with the holonomy sectors will again be carried out correctly
by the methods in this section.

\def\bM{{ \bar{M} }}
\def\a{{ \alpha }}
The doubly twisted
partition function \twisted\ naturally splits into a product of
independent pieces,
only related through the $U(1)$ twists:
\eqn\pfsplit{
Z(\tau,\theta,\bar\theta) = Z_{SU(2)_k} (\tau,\bar\theta)
Z_{SO(2)_1} (\tau,Q\theta,\bar\theta)
Z_{SO(2)_{-{(k+2)}}} (\tau,-\theta,\bar\theta) Z_{ghosts} (\tau) ~.}
In the operator language, these components
can be expressed in terms of characters of integrable representations for the
different affine Lie algebras.

The $SU(2)_k$ WZW model
component consists of a modular invariant combination of
characters
\eqn\cpone{
Z_{SU(2)_k} (\tau,\bar\theta) =
\sum_{L,\bL=0}^k \X_{L;k}(\tau,0) N_{L\bL}
\overline{ \X_{\bL;k}(\tau,\bar\theta) } ~,}
where the integral matrices $N_{L\bL}$ have been classified in
 ref.\ref\ciz{A. Cappelli, C.Itzykson and J-B. Zuber, Commun. Math. Phys.
{\bf 113} (1987)~1.}.
In general, for simply connected groups, the WZW path integral yields the
diagonal modular invariant \gk. We notice, however that the `continuous
orbifold' considerations  should work for more general
invariants. We therefore shall work with more general $N_{L\bL}$ than the
diagonal case in what follows.

The $SO(2)_1$ component is the usual partition function for a
compactified boson of radius one (see e.g.
ref.\ref\pgins{P. Ginsparg, {\sl `Applied Conformal Field Theory'}, Published
in Les Houches Summer School 1988 1--168.)}
\eqn\cptwo{
Z_{SO(2)_1} (\tau,Q\theta,\bar\theta) = \sum_{m,n \in \IZ}
{q^{\half({m \over 2}+n)^2} \e{2i \pi Q\theta ({m \over 2}+n)} \over
\eta(\tau)}
{\bar{q}^{\half({m \over 2}-n)^2} \e{-2i \pi \bar\theta ({m \over 2}-n)}
\over \bar{\eta(\tau)}} ~.}
Finally, the non unitary boson and the ghosts yield a contribution
\eqn\cpthree{
Z_{SO(2)_{-(k+2)}} (\tau,-\theta,\bar\theta) Z_{ghosts} (\tau)
= \sum_{\tm,\tn} {q^{-{\tm^2 \over k+2}} \e{-2i\pi\theta \tm}
\over \eta(\tau)}
{\bar{q}^{-{\tn^2 \over k+2}} \e{-2i\pi\bar\theta \tn} \over \bar{\eta(\tau)}}
| \eta(\tau) |^4 ~.
}
At this stage we do not have enough information about the non unitary theory
to completely specify its spectrum $\tm, \tn$.
We will adopt a two step strategy by
first constraining the representations of the non unitary theory using
the requirement of
modular invariance of the final partition function, and then seeking agreement
with the elliptic genus which we obtained using field theoretical methods
in Section~3.

\subsec{\sl The $(0,2)$ Modular Invariants}
Before performing the integral over the twist, it is convenient to
decompose the affine
$SU(2)$ characters so as to single out their $U(1)$ dependence,
according to
\eqn\sutwospl{
\X^{su(2)}_{L;k}(\tau,\theta)
= \sum_{M=0}^{2k-1} C^L_M(\tau) \Theta_{M;k}(\tau,\theta) ~.
}
The $SU(2)$ level $k$ string functions $C^L_M (\tau)$ are defined in
ref.\ref\kacpet{V.G. Kac and D.H. Peterson, Adv. in Math. 53 (1984) 125.}.
They vanish for $L-M \not\in 2\IZ$. The generalised $\Theta$--functions
are defined as:
\eqn\thetagen{
\Theta_{m;k}(\tau,z) = \sum_{n \in \IZ} \e{2i\pi \tau k (n + m/2k)^2}
\e{2i\pi z k (n + m/2k)}.
}
The twist
integral on the left--moving sector involves only the two
bosons and is particularly simple
\eqn\intlm{
{1 \over 2\pi}
\int d\theta \e{2i\pi \theta \left[Q({m \over 2} + n) - \tm\right]} =
\delta_{Q({m \over 2} + n) - \tm,0} ~.
}
The right--moving sector integral is quite similar, but this time includes a
contribution from the $\Theta$--function in the affine $SU(2)$ character
\eqn\intrm{
{1 \over 2\pi}
\int d\bar\theta \e{-2i\pi \bar\theta \left[k(\bar{p} + {\bM \over 2k}) +
({m \over 2} - n) + \tn\right]} =
\delta_{k(\bar{p} + {\bM \over 2k}) + ({m \over 2} - n) + \tn,0} ~.
}
We choose to solve these two constraints in terms of the integers $m,n$.
This means that the final partition function is non--vanishing only when a
solution exists, that is
when $\tm/Q \in \half\IZ$ and $\tn+\tm/Q+\bM/2 \in 2\IZ$.
Assuming this is true, the partition function~\partend\ is therefore of the
form
\eqn\intermpf{
Z(\tau) = \sum_{\tm,\tn} \sum_{L,\bL=0}^k \sum_{\bM=0}^{2k-1}
\X_L(\tau,0) N_{L\bL}
\bC^{\bL}_{\bM} (\tau) \bar\Theta_{\bar{M}(k+2) + k 2\tn; k(k+2)}
(\tau/2,0) ~.
}
This expression is not too surprising since
we expect to have an affine $SU(2)$
algebra in the left--moving sector, and so we get $SU(2)$ characters. We also
expect an $N=2$ superconformal symmetry for the right--moving sector, where
we see that we have obtained
expressions close to $N=2$ characters\ref\rava{F. Ravanini and S-K. Yang, Phys.
Lett. B 195 (1987) 202.},
were it not for the constraints on $\bM$ deriving from the existence of a
solution to \intlm--\intrm.

Now we can impose the easier requirement of modular invariance, the $T$
invariance. The $q,\bar q$ dependence in \intermpf\ is quickly found to be
\eqn\qdep{
q^{{L(L+2) \over 4(k+2)}-{c\over 24}+\IZ}
\bar{q}^{{\bar{L} (\bar{L}+2) \over 4
(k+2)} - {c \over 24} + {(\tm/Q)^2 - (\tn/Q)^2 \over 2}+\IZ} ~.
}
Since the part of the partition function which reflects the $SU(2)$
symmetry is modular invariant, we can neglect the $L,\bL$ dependence in~\qdep.
Thus, $T$ invariance is satisfied when
\eqn\tinv{
\left({\tm \over Q}\right)^2 - \left({\tn \over Q}\right)^2 \in 2\IZ.}

So in the expression \intermpf,
the variables  which we still need to properly constrain are $\tm,\tn$
but we also have a constraint on
$\bM$ coming from \intrm. Taking the $T$ invariance condition \tinv\ into
account, it is convenient then to introduce new unconstrained integers
$m, a_m, b$ related to the old variables by
\eqn\relate{
{\tm \over Q} = {m \over 2}, \qquad
{\tn \over Q} = {m \over 2}+a_m, \qquad
\bar{M}= 4b - m(1+Q) - 2 a_m Q ~,
}
where for even $m,a_m \in 2\IZ$, and for odd $m, a_m \in \IZ$.
At this stage, this ensures the $T$ invariance of the partition function.
The problem now is the $S$--invariance of
\eqn\thisobj{
\eqalign{
  &Z(\tau) =\cr& \!\!\sum_{m,a_m}\! \sum_{L,\bL=0}^{k}
\!\sum_{b=0}^{Q^2-2}\X_L (\tau) N_{L\bL}\bC^\bL_{4b-m(1+Q)-2a_mQ} (\tau)
\bth_{(4b-m)Q^2-Q(m+2a_m); kQ^2} (\tau)  }}
which we will achieve by carefully choosing the sums over $m,a_m$.

\def\square{\kern1pt\vbox{\hrule height 0.8pt\hbox{\vrule width 0.8pt\hskip 3pt
\vbox{\vskip 6pt}\hskip 3pt\vrule width 0.8pt}\hrule height 0.8pt}\kern1pt}
\def\sq{\mathop{\square}}

As the requirement of $S$ invariance is as usual much more difficult to study,
the details of the computation are in  Appendix~B.
Having imposed  $S$ invariance, we found
several modular invariants which can be written as follows:

\noindent
For $Q$ odd,
\eqn\invonetwo{
\eqalign{
  &Z_1(\tau) = \sum_{L,\bar{L}=0}^k \sum_{b=0}^{Q^2-2} \sum_{m=0}^{Q-1}
\sum_{v=0}^3
\X^{su(2)}_L (\tau) N_{L\bar{L}}
\bar{C}^{\bar{L}}_{4b} (\tau) \bth_{4bQ^2-(4m-v)Q
(Q^2-1);kQ^2} (\tau) ~, \cr
&  Z_2 (\tau) = \sum_{L,\bar{L}=0}^k \sum_{b=0}^{Q^2-2} \sum_{m=0}^{Q-1}
 \sum_{v=0,2}
\X^{su(2)}_L (\tau) N_{L\bar{L}}
\bar{C}^{\bar{L}}_{4b} (\tau) \bth_{4bQ^2-(4m-v)Q
(Q^2-1);kQ^2} (\tau) \cr
   &\,\,\,+ \sum_{v=1,3}
\X^{su(2)}_L (\tau) N_{L\bar{L}}
\bar{C}^{\bar{L}}_{4b-2} (\tau) \bth_{(4b-2)Q^2-(4m-v)Q
(Q^2-1);kQ^2} (\tau) ~,}
}
and for $Q$ even,
\eqn\invpm{\eqalign{
&Z_\mp (\tau)
=\cr &\!\sum_{L,\bar{L}=0}^k\! \sum_{b=0}^{Q^2-2} \!\sum_{m=0}^{Q-1}
\sum_{v=0}^3
\X^{su(2)}_L (\tau) N_{L\bar{L}}
\bar{C}^{\bar{L}}_{4b\mp v} (\tau) \bth_{(4b\mp v)Q^2-(4m-v)Q
(Q^2-1);kQ^2} (\tau) \,.}}
Here $N_{L\bar{L}}$ are all the $SU(2)$ modular
invariants of the $A$ and $D$ series, see equation~$(B.8)$.

\def\X{{\cal X}}
\def\T{{\Theta}}
\def\half{{{1 \over 2}}}
\def\a{{\alpha}}

In order to extract some understanding of the physical content of our
modular invariants, we need to assemble them into a form which facilitates
comparison  with the  $(l,q,s)$ notation familiar in
works on the $(2,2)$ minimal models\gepner.
 We concentrate on the $Q$ even case.
Making use of $k+2=2Q^2$ and (trivially) reversing a sign in $4m-v$ we
  extend the sum on $b$ up to
$k-1$, and change the level for the $\Theta$ functions,  using the
relations:
\eqn\relations{\eqalign{
\T_{m;k}(\tau)= \T_{2m;2k}({\tau \over 2})\,\,\,
{\rm and}\,\,\,
\T_{m;k}({\tau \over 2}) = \T_{m;2k}(\tau)+\T_{m+2k;2k}(\tau)
}.}
After this, we get:
\eqn\intermed{
Z_\pm(\tau) =\sum_{L,\bar L} \sum_{b=0}^{k-1} \sum_{m=0}^{Q-1} \sum_{v=0}^3
\X^{su(2)}_L N_{L \bar L} \bar C^{\bar L}_{4b \pm v} \;
\bar\T_{(4b \pm v)(k+2)-(4m+v)Qk;2k(k+2)}(\tau)
}
We now compare the $\bar C\bar\T$ combination with a character in\gepner:
\eqn\gepchar{
\chi^l_{q;s} = \sum_{b=0}^{k-1} C^l_{4b+q-s} \; \T_{2q+ (4b-s)(k+2);
2k(k+2)}(\tau)
}
An obvious identification with our expressions is
\eqn\qsone{\eqalign{
&q-s=\pm v ~{\rm mod}~2k,\cr
&2q-s(k+2)= \pm v(k+2)-(4m+v)Qk ~{\rm mod}~4k(k+2)
}}
which gives
\eqn\qstwo{
s=vQ\mp v ~{\rm mod}~4 ,  \qquad
q=(4m+v)Q ~{\rm mod}~ 2(k+2)~.
}
Of course, $s$ is only defined modulo 4, since we can always absorb a
factor $4\IZ$ in the $b$ sum. As $Q$ is even, this means that $v=0,2$
describe the Neveu--Schwarz sector and $v=1,3$ the Ramond sector.

We therefore obtain the modular invariants \intermed\
in the $l,q,s$ notation:
\eqn\Partlqs{
Z_\pm(\tau) =\sum_{L,\bar L} \sum_{m=0}^{Q-1} \sum_{v=0}^3
\X^{su(2)}_L N_{L \bar L} \bar \X^{\bar L}_{(4m+v)Q;vQ\mp v}(\tau)
}
Recall that only even $\bar L$ occur in the NS sector,
and only odd $\bar L$ in the R sector.

We have yet to discover the physical content of $Z_\pm$. In general,
the sole requirement of modular invariance does not necessarily select
a combination of characters that represents the partition
function. There are two combinations of boundary conditions that are
modular invariant:
\eqn\twopf{
\hat Z_\pm =\half({\scriptstyle A}\sq_A +{\scriptstyle P}\sq_A +{\scriptstyle
A}\sq_P \pm {\scriptstyle P}\sq_P) ~.  }
We use the convention that the Neveu--Schwarz (Ramond) sector is
labeled by a $A$ ($P$) horizontally; the vertical $P$ denotes the
additional insertion $(-1)^F$ in the trace. In particular, $\hat
Z_\pm$ have the same NS sector, but a different R sector. The same
feature is observed in \Partlqs\ for $Z_\pm$.

The difference $\hat Z_+ - \hat Z_- = {\rm Tr}_R(-1)^F$ is also a
modular invariant, and corresponds to the elliptic genus \ellipic,
with $\gamma_L=0$.  The elliptic genus for our coset models has been
computed in \oohlookone, and it is a straightforward exercise to
verify that
\eqn\itworks{
Z_Q^{(0,2)}(q,0,0) = (-1)^{{Q \over 2}}(Z_-(\tau) - Z_+(\tau))~.
}
The details are contained in Appendix C. This implies that for $Q
\in 4\IZ+2$, the partition function of the coset models is $Z_+(\tau)$
in \Partlqs.

In the calculations above, we have not explicitly taken into account that there
are extra $U(1)$ symmetries which commute with the Hamiltonian. These
correspond to the $U(1)$ of the $(N=2)_R$ and the $U(1)$ of the $SU(2)_L$. We
should label all of our states by their charges under these $U(1)$'s. In order
to discover these labels, we could recompute our partition function as above
but with
particular attention paid to the combinations of currents $(J^3,J,I)$ which are
orthogonal to the gauging currents. These will give rise to the $U(1)$'s we
seek. Alternatively, we can simply note that a bonus of working with the
$\T$--functions is that the $U(1)$ dependence may be extracted at any point due
to the unique extension to generalized $\T$--functions defined for example in
 eqn.\thetagen. This amounts to restoring the familiar $U(1)$ dependence of the
$N=2$ and $SU(2)$ characters in expressions \Partlqs.

\newsec{Some Four Dimensional String Theories}
\subsec{Construction of  Heterotic String Theories}
With the $(0,2)$ minimal models' partition functions in place, we are now
ready to
investigate what we may learn about four dimensional string theories
constructed from them.
As usual, we would like to begin by taking the tensor
product of various copies of
the $(0,2)$ minimal models such that the total (internal) central charge is
equal to 9, on both the left and right.
 With the $(2,2)$ minimal models (choosing the $A$--series) the
number of ways of doing this is 168. With our $(0,2)$ minimal models alone,
there is only one  way! This is due to the fact that the level $k$, given by
 the anomaly equation
$k=2(Q^2-1)=6~,~16~,~30~,~48~,~70~,~96~....$ grows rapidly
because $Q$ is restricted to
being an integer. So we may only construct the $(k=6)^4$ model with four
$(0,2)$ minimal models.
However, it only suffices to include a single $(0,2)$ minimal model among a
product of ordinary $(2,2)$ models to produce a $(0,2)$ $c=9$
compactification, and so the number of such models we can make is
 considerably greater than one, using this procedure of `doping'\foot{The term
is chosen because of the  analogy with similar manipulations performed upon
semiconductor materials to drastically change their  band structure (spectrum)
 for the construction of novel electronic devices.}\
the $(2,2)$ models with $(0,2)$ models.

As the central charges of the left and right parts of the internal theories
are the same, the procedure for constructing a heterotic string theory from
minimal models is much the same as originally presented by Gepner\gepner.
The fact that we have written our lowest lying states in terms of $N=2$,
$(l,q,s)$ indices
makes it straightforward to carry out the two important procedures: Aligning
the boundary conditions in each theory and the generalized GSO projection.
These result in a world sheet  $(N=2)_R$ theory with odd integral right $U(1)$
charges
and hence $N=1$ spacetime supersymmetry, the spacetime supercharge arising
from the worldsheet
$(N=2)_R$ spectral flow operator. We worked in the light cone
 gauge, including therefore the two transverse bosons $\db X^i$ and their
superpartners (which form affine $SO(2)$) on the right, for $c_R=12$, and on
the left the two transverse bosons $\d X^i$ together with the 26 additional
fermions (forming affine $E_8\times SO(10)$) needed for $c_L=24$, giving us a
modular invariant
critical heterotic string theory.

In order to construct a modular invariant partition function for the
tensor of the internal (0,2) minimal models which also preserves $N=2$
super symmetry on the world sheet, we need to align the boundary
conditions of the various theories.  Following the approach in
ref.\ref\gepnertrieste{D. Gepner, in the Proceddings of the
Spring School in Superstrings, Trieste, 1989.},
we have to identify the following components in the
partition function:
\eqn\moreboxes{
NS^+={\scriptstyle A}\sq_A, \qquad NS^-={\scriptstyle P}\sq_A, \qquad
R^+={\scriptstyle A}\sq_P, \qquad R^-= {\scriptstyle P}\sq_P ~.
}

We already know one of them, the elliptic genus $R^-=Z_+-Z_-$ for
$Q \in 4\IZ+2$.  The other Ramond contribution is also easy to compute since
$R^+=(Z_+ +Z_-)_{R}$.
To get the Neveu--Schwarz contributions, we make use of the $S$
modular transformation
\eqn\NSplus{
(Z_+)_{NS} = \half(NS^+ +NS^-) ~\mathop{\rightarrow}^S
{}~\half(NS^+ +R^+)
}
from which we can deduce $NS^+$. $NS^-$ follows either by applying the
$T$ modular transformation on $T:\,NS^+ {\rightarrow} NS^-$, or
as $NS^- = 2 (Z_+)_{NS} - NS^+$.

Performing these operations on \Partlqs, we get the buildings blocks for
the (0,2) minimal models partition function ($Q \in 4\IZ+2$)
\eqn\NSR{
\eqalign{
  NS^\pm = \sum_{even~L} \X_L N_{L \bar L} \sum_{m=0}^{Q-1}& \left(
\bar \X^{\bar L}_{4mQ;0}+\bar \X^{\bar L}_{(4m+2)Q;2}
\pm\bar \X^{\bar L}_{4mQ;2}\pm\bar \X^{\bar L}_{(4m+2)Q;0} \right)\cr
 R^\pm = \sum_{odd~L} \X_L N_{L \bar L} \sum_{m=0}^{Q-1} &\biggl(
\bar \X^{\bar L}_{(4m+1)Q;Q-1}+\bar \X^{\bar L}_{(4m+3)Q;Q-3}\cr
&\qquad\qquad\pm\bar \X^{\bar L}_{(4m+1)Q;Q-3}\pm\bar \X^{\bar L}_{(4m+3)Q;Q-1}
\biggr)\cr
}}
{}From these expressions, the product partition function  can be constructed as
\eqn\total{Z_{\rm prod}=\half\left(
\prod_i NS_i^++\prod_i NS_i^-+\prod_i R_i^++\prod_i R_i^-\right).}
The product is over each component theory, including the $SO(2)$
(spacetime fermions) and  $SO(10)\times E_8$ (current algebra fermions)
contributions from the right and left respectively\foot{The $E_8$ sector will
only ever contribute a singlet here, playing its customary
role as the `hidden' gauge
sector.}: \eqn\contributions{NS^\pm_{g/ST} = (C_0 \pm C_2)(\bar B_0 \pm \bar
B_2), \qquad
R^\pm_{g/ST} = (C_1 \pm C_3)(\bar B_1 \pm \bar B_3).
}
 where $C_v$ stands
for the character of the $SO(10)\times E_8$ representation $C_v=(-1)^v
D_v {\bf 1}_{E_8}$. $D_v$ is a $SO(10)$ representation labeled
by $(v)=({\bf 10},{\bf 16},{\bf 1},{\bf \overline{16}})$. $B_v$ is the
character of the $SO(2)$ representation $(v)=({\bf 1},{\bf s},{\bf
v},{\bf \bar s})$.

After a little bit of algebra, the partition function can be expressed as
(neglecting as usual the transverse bosons)
\eqn\expressyourself{\eqalign{
Z_{\rm prod} = \sum_{v=0}^{3} \sum_{l_i,\bar l_i=0}^k \sum_{d_i=0}^{Q-1}
\sum_{r_i,s_i \equiv_2 v} C_v \overline{B}_{v+\sum r_i+\sum s_i}\!\prod_i
\!
\X_{l_i}^{su(2)} N_{l_i,\bar l_i}
\overline{\X}^{\bar l_i}_{(4d_i+r_i)Q;r_i Q-s_i}.
}}

The next step in constructing the heterotic string theory is to realize
spacetime supersymmetry. This is done by projecting out all states which have
other than odd integer right $U(1)$ charge. Of course, this  projection must be
done in a modular invariant way, requiring the inclusion of  twisted sectors.
In particular this will  build for us in spacetime the  spin 3/2 gravitino, and
 simultaneously remove tachyons from the physical spectrum, leaving tachyons
to
contribute to the string theory only in loop amplitudes.

The three phenomena of modular invariance, odd--integer $U(1)_R$ and spacetime
supersymmetry are all crucially interlinked, of course, and are implemented
by  the familiar {
GSO projection}. Under modular transformations, the  characters of a
 spacetime supersymmetric model transform into sums of characters of a
worldsheet
 $N=2$ theory with only odd integrally charged $U(1)_R$ states present.
This forms  a  unitary representation of the modular group, as can be checked
explicitly.
The most efficient way to carry out this orbifolding procedure is to work
directly with the  world sheet  $N=2$ spectral flow operator for the  $c_R=12$
system, noting that under its action relating the $NS$ and $R$ sectors, the
$U(1)$ charge changes by $c_R/6=2$. Starting with the modular invariant
non--spacetime--supersymmetric partition function,
one simply needs to generate all of the  states which can be reached by the
action of the  flow operator (the twisted sectors of the orbifold)  and project
according to the charge condition in each sector. The fact that the right
$U(1)$ charge changes by two under the action of the  spectral flow guarantees
that action of the spacetime supercharge is well defined on states in the
theory.

The  gauge symmetry arising from these models will arise from the left as usual
as
$E_8\times SO(10)\times {\widetilde G}$.
All states in the model are singlets under the $E_8$, while
  $\widetilde G$ is an  enhanced gauge symmetry arising from the affine
structures in the left part of the internal theory. Vertex operators for
creation of spacetime vectors corresponding to gauge bosons of $\widetilde G$
can be
constructed as  \eqn\vertex{V^{\mu a}=<1|<\!J^a_{-1}\cdot  1||\psi^\mu\!>|1>.}
The first contribution is the singlet from $SO(10)$, the second is a descendent
of the vacuum of the internal theory under the affine current mode $J^a_{-1}$.
This gives a state of left conformal weight 1.
The third contribution $\psi^\mu$ is the $SO(2)$ vector with right conformal
weight 1/2
and charge 1, and the fourth is the $NS$ vacuum of the internal theory.
 The operator $V^{\mu a}$ is thus an allowed massless vector in the
theory, as $c_R=12$ and $c_L=24$. The action of spectral flow will fill out the
enhanced
gauge supermultiplet.
  Each constituent of the internal theory has either a $U(1)$ affine symmetry
(if it is a $(2,2)$ minimal model) or an $SU(2)_k$ (if it is a $(0,2)$
 minimal model at level $k$).
Thus, the enhanced gauge symmetry group $\widetilde G$ is a product of factors
made up of $U(1)$'s and $SU(2)_k$'s. There is also the possibility of
`accidental' contributions to the enhanced gauge symmetry group
 occurring when there is a means
of constructing  a vertex operator for massless $SO(10)$ singlets which are
spacetime vectors, as above, but now the descendants $<\!J^a_{-1}\cdot 1|$ are
replaced instead by weight 1 states coming from the internal sector which are
{\sl
not} descendents. We shall see this occurring in one of our examples.

\subsec{Moving to $E_6$ Gauge Symmetry}
There would seem to be  the possibility that for a particular
 model  with $c_{L~{\rm int}}=c_{R~{\rm int}}=9$ the
$SO(10)$ representations and representations of a diagonal  $U(1)$ subgroup of
$\widetilde G$
might fill
 out complete $E_6$ representations, as happens for the $(2,2)$ models.
 For this to happen there must exist
 an operator in the theory on the left which
acts as a `spectral flow' operator, this time relating the
 various $SO(10)\times U(1)$ representations. The presence of such an operator
is of course guaranteed in the case where the internal theory was built out of
a $(2,2)$
model, as the internal $(N=2)_L$ has a spectral flow operator from which
such an object is built. (Indeed, the analogue of the gravitino in this case is
the gaugino  transforming as the anti--spinor
$\bar {\bf 16}$ of $SO(10)$ with internal $U(1)$ charge  $3/2$.) In the realm
of $(0,2)$ models, it is not necessary that such an operator exists, as we see
in half of the examples we present below\foot{There are  models in the
literature where such an operator is present, however.~See
for example
refs.\ref\jacques{J. Distler, in the proceedings of the Trieste Summer School
on High Energy Physics, Trieste, 1994, hep-th/9502012.}\dkone\dktwo\cfttwo.}.
However, it is interesting to see that one can construct such an operator and
use it to enhance $SO(10)$ times a diagonal $U(1)$ to $E_6$ if one so desired.
One can choose a suitably normalised $U(1)$ subgroup of $\widetilde G$ for this
purpose,  normalising the currents $J(z)$ such that
$J(z)J(w)\sim 3/(z-w)^{-2}$.  By bosonising this current according to
$J=i\sqrt{3}\partial\varphi$ one can  rewrite all fields $f_q$ in the theory
with charge $q$
under this $U(1)$ as the product $f_q=f\cdot\exp{(iq\varphi/\sqrt{3})}$.
The spectral flow operator we need is simply the action of
$Q_{\rm int}=\exp{(i\sqrt{3}\varphi/2)}$ in the internal sector which has
conformal dimension 3/8 and charge $3/2$. The total (weight 1) spectral flow
operator for the left is made by multiplying  this by the $\bar {\bf 16}$ from
the SO(10) theory, which has weight 5/8 and $U(1)$ charge  $1/2$. This state,
the
${\bar {\bf 16}}_{3/2}$, and its conjugate ${ {\bf 16}}_{-3/2}$ (after
appropriate dressing from the right states) forms part of the gauge
supermultiplet
of $E_6$ via the decomposition
${\bf 78}= {\bf 1}_{0}+{\bar {\bf 16}}_{3/2}+{\bf 16}_{-3/2}+{\bf 45}_0$,
where {\bf 45} and {\bf 1} denote the adjoint and singlet of $SO(10)$.
 Notice that the action of the total spectral flow operator again changes the
total $U(1)$
charge of a state by 2. To arrive at an $E_6$ model we can simply construct
this operator and use it to project  onto even\foot{Projecting onto even
integer of course, as we would like to retain  for example the graviton, which
has charge zero. Note that this projection is automatic in the $(2,2)$ case for
the reasons described in the text.}\ integer $U(1)_L$ in an analogous
procedure to that carried out on the right for spacetime supersymmetry.

In the examples we study here, we use our $SO(10)\times {\widetilde
G}$ models as a starting point. Each $SU(2)_6$ constituent
of the left part of the internal theory has a $U(1)$ subgroup which
contributes to the left spectral flow operator. Acting with the
operator on the $SU(2)$ pieces has the effect of isolating the
parafermion piece while modifying the $U(1)$ contribution from which
$SU(2)$ is made. The total internal $U(1)$ current is given by $J_{\rm
int}=\prod_iJ_i+1/2\prod_iJ^3_i$, where the $(2,2)$ minimal models
each contribute a $J$ and the $(0,2)$ minimal models each supply a
$J^3$ from their Cartan subalgebra.  The result of twisting the
$SU(2)$ contribution from a $(0,2)$ minimal model can be summarised
succinctly by decomposing the usual character \sutwospl\ into (recall
that we use $J^3/2$ for the $U(1)$)
\eqn\twistedsutwo{
\X_{l;k}(\tau,{z \over 2})
= \sum_{m=0}^{4k-1} C^l_m(\tau) \Theta_{2m;4k}(\tau,{z\over4}) =
\sum_{m=0}^{4k-1} Y^l_{m;n=2m}(\tau,z).
}
This is similar to the decomposition of a full $N=2$ character into
$\X^l_{q;s}$ functions. States described by the $Y^l_{m;n}$ have
dimensions and charges
\eqn\eigen{
\eqalign{
\Delta &= {l(l+2) \over 4(k+2)} - {m^2 \over 4k} + {n^2 \over 16k} + \IZ \cr
Q &= {n \over 8} + 2\IZ}
}
(The familiar $SU(2)$ states are those for $n=2m$). The action of the twist
amounts simply to $Y^l_{m;n} \rightarrow Y^l_{m;n+3}$.
It is easy to check that this twist (flow) action is to change
 the dimension and charge of a state by $q/2+c/24$  and $c/6$ respectively, for
each model, where $c=9/4$, which is precisely the same as for the action of a
spectral flow  operator on an $N=2$ state. Therefore for the action of the
total flow  operator, the $U(1)$ charge of this non--supersymmetric
internal theory is again changed by $3/2$ as for an
$N=2$ supersymmetric model!
Once we combine this operator   with the current algebra sector, its action
will again change the total $U(1)$ charge by 2.
Using this flow  operator, we can obtain modular invariant partition functions
for heterotic string theory with a linearly realised\foot{This of course will
readily generalise to the case when the central charges of the internal
theory are not equal, giving linear realisations of $SO(10)$ and  $SU(5)$, for
example\jacques.}\ $E_6$ by a constructive method exactly analogous to that
described above to realise spacetime supersymmetric models
using the  spectral flow operator of the right $N=2$.

The result of projecting onto even $U(1)$ charge
is that the $SU(2)$ factors from each theory get broken  to $U(1)$. This is
easy to see, as the descendents (discussed above) which make their gauge bosons
all now have the wrong charge under the projecting $U(1)$. All that is, except
the abelian contribution. In this way, we constructed a family of
$E_6\times U(1)^3$ models, presented below. This procedure, like any of those
described above, is easily generalised to different situations, for example
the numerous `doped' models which can be made.

We now go on to describe the models which we studied using all of the
 methods we described above.

\subsec{Four $SO(10)\times\widetilde G$ examples}
Let us start by considering the  string theory where all the four
factors in the internal theory are $A$--series
$(0,2)$ minimal models at $k=6$.
We shall call this the (0000) model\foot{Our notation
shall be of the form (WXYZ) where a letter  is a `0' for a $(0,2)$ minimal
model and a `2' for a $(2,2)$ minimal model. The pattern of constituent
models can thus be easily  read. We shall  use $(WXYZ)_{E_6}$ to refer to the
models which have $E_6$ gauge symmetry.}.
The computation of the massless spectrum
 is obtained by the application of the procedures described above and
the results of which can be found below. (For simplicity, in listing the
massless matter content of a model  we shall mention only  the
number of (spacetime) scalars in each sector, specifically
those  which are the superpartners
of (spacetime) right--moving fermions. $N=1$ spacetime supersymmetry  and
CPT invariance are of course present in these consistent models, and therefore
the reader may deduce the rest of the content of massless matter
sector---superpartners and antiparticles---at their leisure. Consulting tables
5.1 will yield more
 details of the spectrum in each example.)

\noindent
So for the (0000) model we have
\item{$\bullet$} 4 scalars in the ${\bf 16}$ $SO(10)$.
\item{$\bullet$} 6  scalars
in the $\bf 10$ of $SO(10)$.
\item{$\bullet$} 13 scalars which are singlets of $SO(10)$.

\noindent
In addition there are the gauge degrees of freedom corresponding to the
gauge symmetry $SO(10)\times SU(2)^4$.

Let us consider the $SO(10)$
singlets. There are 13 scalars, together with their
fermionic superpartners and all  of  the antiparticles.  An interesting
and important
question is whether any of them are moduli, as with this knowledge we may begin
to understand if there is any way to reach a sigma--model with spacetime
geometrical interpretation from  these models. To answer this,
we need to check for  the existence of exactly flat directions. As we know
the partition function we could in principle compute various
correlation functions to establish the presence of such exactly marginal
deformations. However, the existence of an
R--symmetry, the quantum symmetry appearing in the GSO projection in
labeling the twisted sectors, sometimes
helps us to argue for flatness of untwisted
singlets\foot{This
procedure has been used in arguing the flatness of  $E_6$ singlets in
certain Landau-Ginzburg orbifolds~\dktwo.}.
Following~\ref\kw{S.~Kachru and E.~Witten, \npb{407}{1993}{637},
{ hep-th/9307038}.}\ we have
$
R=\exp(2\pi iN/4)
$
where  the order of the discrete symmetry\foot{Note that for
 the `doped' models it is the more
familiar $2(k+2)=16$. While a single $(2,2)$ minimal model has a
$\IZ_{k+2}\times \IZ_2$
symmetry associated with it, a single $(0,2)$ minimal model has $\IZ_{2Q}$.}\
 associated to the
GSO projection is in this case, 4.
As the spacetime superpotential transforms with charge $-2$ mod
$16$ we need to make sure that there are no couplings involving the
$SO(10)$ singlets $\Phi_i, i=1,...,13$ of the form
$f(\Phi_i)\Phi_{j\neq i}^{s}$ for $s=0,1$ where $f(\Phi_{i})$ is
a function
of untwisted singlets only while $\Phi_{j}$ is
 any $SO(10)$ singlet. For this example, there is one
singlet which  has charge $-2$ {\rm mod}
$4$, and its presence thus prevents us from arguing for
 the presence of flat directions preserving $SO(10)$.
 However, by examining the  charges of all of the fields
under  the enhanced gauge symmetry $SU(2)^4$, we see that there are
 6 $SO(10)\times SU(2)$ singlets from the untwisted sector
which cannot  form couplings in the superpotential which
would spoil flatness.
So we have six moduli which preserve $SO(10)\times SU(2)$.
Further analyses of this type, using R--symmetry and extended gauge symmetry
may reveal further flat directions.
It should be noted however, that once these methods are exhausted,
 it could well be
 that the possibility
of couplings between $SO(10)$ singlets that would ruin flatness
does not happen, simply because the couplings vanish of their own accord
due to details of the conformal field theory.
More work is needed to determine whether this is indeed the case.
See table 5.1a for a summary of how the scalar singlets fall into the twisted
sectors. (Note that the occupancy of the other twisted sectors can be deduced
from supersymmetry.)

Although it was pointed out above there is only one choice of a
$(k=6)^4$ tensor product of purely $(0,2)$ minimal model
factors, we have the
possibility of replacing one (or more) of these by the standard
$(2,2)$ minimal models, a procedure we referred to as `doping'.
 The result of all these
computations, carried out in the same fashion as above, can be found in
tables~5.1b--d at the end of this section. Notice that the number of
singlets increases with the amount of doping by $(2,2)$ minimal models,
as does the number of $\bf 16$'s and $\bf 10$'s. In all cases, there are
singlets of charge $-2$ under the $\IZ_{16}$
R--symmetry, but upon requiring preservation of
some of the enhanced gauge symmetry, it is readily
seen that at least some of those from the untwisted sector  are moduli.

In addition there are massless vectors  corresponding to the
gauge symmetry
 $SO(10)\times SU(2)^2\times U(1)^2$ for the $(2200)$ model and
$SO(10)\times SU(2)\times U(1)^3$ for the $(2220)$ model.
In the case of the $(2000)$ model, however, the gauge symmetry is
 $SO(10)\times SU(2)^3\times U(1)^3$. The two extra `accidental'
$U(1)$'s arise
because there are two additional (1,0) currents in this  model which are not
related to descendent states under the affine symmetry of any of the individual
models.

\subsec{Four $E_6\times U(1)^3$ examples}
As described above, we took the models of the previous section and constructed
four $E_6$ $(0,2)$  string vacua from  them, by acting with a left spectral
flow  operator. The particle content of these models is listed in table
5.2.
The
$(0000)_{E_6}$ model, which  has a $\IZ_4$
R-symmetry, has

\item{$\bullet$} 23 scalars in the ${\bf 27}$ $E_6$.
\item{$\bullet$} 7  scalars
in the $\bar{\bf 27}$ of $E_6$.
\item{$\bullet$} 191 scalars which are singlets of $E_6$.

For this model   there are
 singlets of charge $-2$ under the $\IZ_4$
R--symmetry requiring us to find more powerful methods of
determining the moduli, as discussed in the previous subsection.

Turning to the `doped' models, we find a number of interesting facts. First,
the $(2000)_{E_6}$ model has the {\sl same spectrum} as the
$(0000)_{E_6}$ model! The projection to realise $E_6$  has resulted in two
equivalent models, thereby negating the effect of the doping.  Examining the
details of the spectrum, we see that  $(2000)_{E_6}$ has naively a
$\IZ_{16}$ R--symmetry. However, if we relabel the $N$th twisted sector as $N$
mod 4, thus realising an $\IZ_4$ symmetry, then the two spectra are indeed
identical.

The $(2200)_{E_6}$ model, which has $\IZ_{16}$ R--symmetry, has

\item{$\bullet$} 49 scalars in the ${\bf 27}$ $E_6$.
\item{$\bullet$} 9  scalars
in the $\bar{\bf 27}$ of $E_6$.
\item{$\bullet$} 251 scalars which are  $E_6$ singlets.

Again, there are singlets of charge $-2$ under the R--symmetry.

The $(2220)_{E_6}$ model, which  also has $\IZ_{16}$ R--symmetry, has
a spectrum which is {\sl identical} to that of the familiar
$(2222)_{E_6}$ $(2,2)$ Gepner model! This time, the projection realising $E_6$
has negated the  doping.
Indeed, upon
examination of the  explicit form of the partition function in terms of the
various characters, (appropriately projected)
 we have been able to show that
they are indeed equivalent models.
The spectrum is:

\item{$\bullet$} 149 scalars in the ${\bf 27}$ $E_6$.
\item{$\bullet$} 1  scalar
in the $\bar{\bf 27}$ of $E_6$.
\item{$\bullet$} 503  scalars which are singlets under  $E_6$.

 There are some general features to remark upon.
In all cases
the gauge group is  $E_6\times U(1)^3$, the $SU(2)$ factors having been
broken by the projection. Further to this is the increase with the degree of
(effective) doping  in the net number of
chiral generations (number of $\bf 27$'s minus number of $\bar{\bf 27}$'s)
(16, 40, 148) for $(0,2000)_{E_6}$, $(2200)_{E_6}$ and  $(2220,2)_{E_6}$,
respectively. Also increasing is the number of singlets, (191,251,503).
In addition,
the number of singlets
in the untwisted sector keeps increasing with respect to the number in any
other
single sector.

Whether these features will persist in other doping examples is not known. It
is also not known whether they are of any significance.  These questions will
probably be answered when methods for determining the full set of  moduli
of these models are uncovered. The unexpected fact that in these examples,
doping with either just one $(0,2)$ or just one $(2,2)$ minimal
model has no effect after projecting to recover $E_6$ is interesting.
This feature appears to be present in other examples. However, it is not known
what the general pattern is when there is more than one (0,2) factor present,
or if the levels are not all the same. Further investigation into these matters
is continuing.

\medskip

\vbox{
$$\vbox{\offinterlineskip
\hrule height 1.1pt
\halign{&\vrule width 1.1pt#&\strut\quad#\hfil\quad&
\vrule#&\strut\quad#\hfil\quad&
\vrule#&\strut\quad#\hfil\quad&
\vrule#&\strut\quad#\hfil\quad&
\vrule#&\strut\quad#\hfil\quad&
\vrule#&\strut\quad#\hfil\quad&\vrule width 1.1pt#\cr
height3pt&\omit&&\omit&&\omit&&\omit&\cr
&\hfil$N$&&\hfil ${\bf 10}$&&\hfil ${\bf 16}$&&${\bf 1}$&\cr
height3pt&\omit&&\omit&&\omit&&\omit&\cr
\noalign{\hrule height 1.1pt\vskip3pt\hrule height 1.1pt}
height3pt&\omit&&\omit&&\omit&&\omit&\cr
&0&
&6&&&&12&\cr
height3pt&\omit&&\omit&&\omit&&\omit&\cr
\noalign{\hrule}
height3pt&\omit&&\omit&&\omit&&\omit&\cr
&1&
&&&&&&\cr
height3pt&\omit&&\omit&&\omit&&\omit&\cr
\noalign{\hrule}
height3pt&\omit&&\omit&&\omit&&\omit&\cr
&2&
&&&&&1&\cr
height3pt&\omit&&\omit&&\omit&&\omit&\cr
\noalign{\hrule}
height3pt&\omit&&\omit&&\omit&&\omit&\cr
&3&
&&&4&&&\cr
height3pt&\omit&&\omit&&\omit&&\omit&\cr}
\hrule height 1.1pt}
$$
\vskip5pt
\noindent
{\bf Table 5.1a}: The spectrum for the massless matter fields
(spacetime scalars which are lowest components of
chiral superfields) in the
$(0000)$ model, in $N$th twisted sector of the GSO projection.
There are 6 ${\bf 16}$'s, 4 ${\bf 10}$'s and 13 ${\bf 1}$'s of $SO(10)$
respectively. The corresponding superpartners and antiparticles
are obtained by spectral flow  and CPT--conjugation respectively.}

\medskip

\vbox{
$$\vbox{\offinterlineskip
\hrule height 1.1pt
\halign{&\vrule width 1.1pt#&\strut\quad#\hfil\quad&
\vrule#&\strut\quad#\hfil\quad&
\vrule#&\strut\quad#\hfil\quad&
\vrule#&\strut\quad#\hfil\quad&
\vrule#&\strut\quad#\hfil\quad&
\vrule#&\strut\quad#\hfil\quad&\vrule width 1.1pt#\cr
height3pt&\omit&&\omit&&\omit&&\omit&\cr
&\hfil$N$&&\hfil ${\bf 10}$&&\hfil ${\bf {16}}$&&${\bf 1}$&\cr
height3pt&\omit&&\omit&&\omit&&\omit&\cr
\noalign{\hrule height 1.1pt\vskip3pt\hrule height 1.1pt}
height3pt&\omit&&\omit&&\omit&&\omit&\cr
&0&
&6&&&&12&\cr
height3pt&\omit&&\omit&&\omit&&\omit&\cr
\noalign{\hrule}
height3pt&\omit&&\omit&&\omit&&\omit&\cr
&2&
&&&&&1&\cr
height3pt&\omit&&\omit&&\omit&&\omit&\cr
\noalign{\hrule}
height3pt&\omit&&\omit&&\omit&&\omit&\cr
&3&
&&&3&&&\cr
height3pt&\omit&&\omit&&\omit&&\omit&\cr
\noalign{\hrule}
height3pt&\omit&&\omit&&\omit&&\omit&\cr
&4&
&&&&&3&\cr
height3pt&\omit&&\omit&&\omit&&\omit&\cr
\noalign{\hrule}
height3pt&\omit&&\omit&&\omit&&\omit&\cr
&6&
&&&&&1&\cr
height3pt&\omit&&\omit&&\omit&&\omit&\cr
\noalign{\hrule}
height3pt&\omit&&\omit&&\omit&&\omit&\cr
&8&
&3&&&&6&\cr
height3pt&\omit&&\omit&&\omit&&\omit&\cr
\noalign{\hrule}
height3pt&\omit&&\omit&&\omit&&\omit&\cr
&11&
&&&1&&&\cr
height3pt&\omit&&\omit&&\omit&&\omit&\cr
\noalign{\hrule}
height3pt&\omit&&\omit&&\omit&&\omit&\cr
&12&
&3&&&&6&\cr
height3pt&\omit&&\omit&&\omit&&\omit&\cr
\noalign{\hrule}
height3pt&\omit&&\omit&&\omit&&\omit&\cr
&14&
&&&&&1&\cr
height3pt&\omit&&\omit&&\omit&&\omit&\cr
\noalign{\hrule}
height3pt&\omit&&\omit&&\omit&&\omit&\cr
&15&
&&&4&&&\cr
height3pt&\omit&&\omit&&\omit&&\omit&\cr}
\hrule height 1.1pt}
$$
\vskip5pt
\noindent
{\bf Table 5.1b}: The spectrum for the massless matter fields
in the $(2000)$ model.
There are 12 ${\bf 16}$'s, 8 ${\bf 10}$'s and 30 ${\bf 1}$'s of $SO(10)$
respectively.}

\vbox{
$$\vbox{\offinterlineskip
\hrule height 1.1pt
\halign{&\vrule width 1.1pt#&\strut\quad#\hfil\quad&
\vrule#&\strut\quad#\hfil\quad&
\vrule#&\strut\quad#\hfil\quad&
\vrule#&\strut\quad#\hfil\quad&
\vrule#&\strut\quad#\hfil\quad&
\vrule#&\strut\quad#\hfil\quad&\vrule width 1.1pt#\cr
height3pt&\omit&&\omit&&\omit&&\omit&\cr
&\hfil$N$&&\hfil ${\bf 10}$&&\hfil ${\bf {16}}$&&${\bf 1}$&\cr
height3pt&\omit&&\omit&&\omit&&\omit&\cr
\noalign{\hrule height 1.1pt\vskip3pt\hrule height 1.1pt}
height3pt&\omit&&\omit&&\omit&&\omit&\cr
&0&
&16&&&&38&\cr
height3pt&\omit&&\omit&&\omit&&\omit&\cr
\noalign{\hrule}
height3pt&\omit&&\omit&&\omit&&\omit&\cr
&2&
&&&&&5&\cr
height3pt&\omit&&\omit&&\omit&&\omit&\cr
\noalign{\hrule}
height3pt&\omit&&\omit&&\omit&&\omit&\cr
&3&
&&&2&&&\cr
height3pt&\omit&&\omit&&\omit&&\omit&\cr
\noalign{\hrule}
height3pt&\omit&&\omit&&\omit&&\omit&\cr
&4&
&2&&&&10&\cr
height3pt&\omit&&\omit&&\omit&&\omit&\cr
\noalign{\hrule}
height3pt&\omit&&\omit&&\omit&&\omit&\cr
&6&
&&&&&1&\cr
height3pt&\omit&&\omit&&\omit&&\omit&\cr
\noalign{\hrule}
height3pt&\omit&&\omit&&\omit&&\omit&\cr
&7&
&&&1&&&\cr
height3pt&\omit&&\omit&&\omit&&\omit&\cr
\noalign{\hrule}
height3pt&\omit&&\omit&&\omit&&\omit&\cr
&8&
&1&&&&4&\cr
height3pt&\omit&&\omit&&\omit&&\omit&\cr
\noalign{\hrule}
height3pt&\omit&&\omit&&\omit&&\omit&\cr
&12&
&3&&&&8&\cr
height3pt&\omit&&\omit&&\omit&&\omit&\cr
\noalign{\hrule}
height3pt&\omit&&\omit&&\omit&&\omit&\cr
&14&
&&&&&7&\cr
height3pt&\omit&&\omit&&\omit&&\omit&\cr
\noalign{\hrule}
height3pt&\omit&&\omit&&\omit&&\omit&\cr
&15&
&&&13&&&\cr
height3pt&\omit&&\omit&&\omit&&\omit&\cr}
\hrule height 1.1pt}
$$
\vskip5pt
\noindent
{\bf Table 5.1c}: The spectrum for the massless matter fields
 in the $(2200)$ model.
There are 22 ${\bf 16}$'s, 16 ${\bf 10}$'s and 73 ${\bf 1}$'s of $SO(10)$
respectively.}

\medskip

\vbox{
$$\vbox{\offinterlineskip
\hrule height 1.1pt
\halign{&\vrule width 1.1pt#&\strut\quad#\hfil\quad&
\vrule#&\strut\quad#\hfil\quad&
\vrule#&\strut\quad#\hfil\quad&
\vrule#&\strut\quad#\hfil\quad&
\vrule#&\strut\quad#\hfil\quad&
\vrule#&\strut\quad#\hfil\quad&\vrule width 1.1pt#\cr
height3pt&\omit&&\omit&&\omit&&\omit&\cr
&\hfil$N$&&\hfil ${\bf 10}$&&\hfil ${\bf {16}}$&&${\bf 1}$&\cr
height3pt&\omit&&\omit&&\omit&&\omit&\cr
\noalign{\hrule height 1.1pt\vskip3pt\hrule height 1.1pt}
height3pt&\omit&&\omit&&\omit&&\omit&\cr
&0&
&51&&&&138&\cr
height3pt&\omit&&\omit&&\omit&&\omit&\cr
\noalign{\hrule}
height3pt&\omit&&\omit&&\omit&&\omit&\cr
&2&
&3&&&&22&\cr
height3pt&\omit&&\omit&&\omit&&\omit&\cr
\noalign{\hrule}
height3pt&\omit&&\omit&&\omit&&\omit&\cr
&3&
&&&1&&&\cr
height3pt&\omit&&\omit&&\omit&&\omit&\cr
\noalign{\hrule}
height3pt&\omit&&\omit&&\omit&&\omit&\cr
&4&
&&&&&9&\cr
height3pt&\omit&&\omit&&\omit&&\omit&\cr
\noalign{\hrule}
height3pt&\omit&&\omit&&\omit&&\omit&\cr
&6&
&&&&&1&\cr
height3pt&\omit&&\omit&&\omit&&\omit&\cr
\noalign{\hrule}
height3pt&\omit&&\omit&&\omit&&\omit&\cr
&8&
&&&&&3&\cr
height3pt&\omit&&\omit&&\omit&&\omit&\cr
\noalign{\hrule}
height3pt&\omit&&\omit&&\omit&&\omit&\cr
&14&
&&&&&34&\cr
height3pt&\omit&&\omit&&\omit&&\omit&\cr
\noalign{\hrule}
height3pt&\omit&&\omit&&\omit&&\omit&\cr
&15&
&&&43&&&\cr
height3pt&\omit&&\omit&&\omit&&\omit&\cr}
\hrule height 1.1pt}
$$
\vskip5pt
\noindent
{\bf Table 5.1d}: The spectrum for the massless matter fields
 in the $(2220)$ model.
There are 54 ${\bf 16}$'s, 44 ${\bf 10}$'s and 207 ${\bf 1}$'s of $SO(10)$
respectively.}

\medskip

\vbox{
$$\vbox{\offinterlineskip
\hrule height 1.1pt
\halign{&\vrule width 1.1pt#&\strut\quad#\hfil\quad&
\vrule#&\strut\quad#\hfil\quad&
\vrule#&\strut\quad#\hfil\quad&
\vrule#&\strut\quad#\hfil\quad&
\vrule#&\strut\quad#\hfil\quad&
\vrule#&\strut\quad#\hfil\quad&\vrule width 1.1pt#\cr
height3pt&\omit&&\omit&&\omit&&\omit&\cr
&\hfil$N$&&\hfil ${\bf 27}$&&\hfil ${\bf \bar{27}}$&&${\bf 1}$&\cr
height3pt&\omit&&\omit&&\omit&&\omit&\cr
\noalign{\hrule height 1.1pt\vskip3pt\hrule height 1.1pt}
height3pt&\omit&&\omit&&\omit&&\omit&\cr
&0&
&6&&6&&60&\cr
height3pt&\omit&&\omit&&\omit&&\omit&\cr
\noalign{\hrule}
height3pt&\omit&&\omit&&\omit&&\omit&\cr
&1&
&&&&&&\cr
height3pt&\omit&&\omit&&\omit&&\omit&\cr
\noalign{\hrule}
height3pt&\omit&&\omit&&\omit&&\omit&\cr
&2&
&1&&1&&51&\cr
height3pt&\omit&&\omit&&\omit&&\omit&\cr
\noalign{\hrule}
height3pt&\omit&&\omit&&\omit&&\omit&\cr
&3&
&16&&&&80&\cr
height3pt&\omit&&\omit&&\omit&&\omit&\cr}
\hrule height 1.1pt}
$$
\vskip5pt
\noindent
{\bf Table 5.2a}: The spectrum for the massless matter fields
 in the
$(0000)_{E_6}$ model.
There are 23 ${\bf 27}$'s, 7 ${\bf \bar {27}}$'s and 191 ${\bf 1}$'s of $E_6$
respectively.}

\medskip

\vbox{
$$\vbox{\offinterlineskip
\hrule height 1.1pt
\halign{&\vrule width 1.1pt#&\strut\quad#\hfil\quad&
\vrule#&\strut\quad#\hfil\quad&
\vrule#&\strut\quad#\hfil\quad&
\vrule#&\strut\quad#\hfil\quad&
\vrule#&\strut\quad#\hfil\quad&
\vrule#&\strut\quad#\hfil\quad&\vrule width 1.1pt#\cr
height3pt&\omit&&\omit&&\omit&&\omit&\cr
&\hfil$N$&&\hfil ${\bf 27}$&&\hfil ${\bf \bar{27}}$&&${\bf 1}$&\cr
height3pt&\omit&&\omit&&\omit&&\omit&\cr
\noalign{\hrule height 1.1pt\vskip3pt\hrule height 1.1pt}
height3pt&\omit&&\omit&&\omit&&\omit&\cr
&0&
&6&&&&36&\cr
height3pt&\omit&&\omit&&\omit&&\omit&\cr
\noalign{\hrule}
height3pt&\omit&&\omit&&\omit&&\omit&\cr
&2&
&1&&&&20&\cr
height3pt&\omit&&\omit&&\omit&&\omit&\cr
\noalign{\hrule}
height3pt&\omit&&\omit&&\omit&&\omit&\cr
&3&
&3&&&&50&\cr
height3pt&\omit&&\omit&&\omit&&\omit&\cr
\noalign{\hrule}
height3pt&\omit&&\omit&&\omit&&\omit&\cr
&4&
&&&3&&9&\cr
height3pt&\omit&&\omit&&\omit&&\omit&\cr
\noalign{\hrule}
height3pt&\omit&&\omit&&\omit&&\omit&\cr
&6&
&&&1&&19&\cr
height3pt&\omit&&\omit&&\omit&&\omit&\cr
\noalign{\hrule}
height3pt&\omit&&\omit&&\omit&&\omit&\cr
&7&
&&&&&15&\cr
height3pt&\omit&&\omit&&\omit&&\omit&\cr
\noalign{\hrule}
height3pt&\omit&&\omit&&\omit&&\omit&\cr
&8&
&&&&&6&\cr
height3pt&\omit&&\omit&&\omit&&\omit&\cr
\noalign{\hrule}
height3pt&\omit&&\omit&&\omit&&\omit&\cr
&10&
&&&&&3&\cr
height3pt&\omit&&\omit&&\omit&&\omit&\cr
\noalign{\hrule}
height3pt&\omit&&\omit&&\omit&&\omit&\cr
&11&
&3&&&&&\cr
height3pt&\omit&&\omit&&\omit&&\omit&\cr
\noalign{\hrule}
height3pt&\omit&&\omit&&\omit&&\omit&\cr
&12&
&&&3&&9&\cr
height3pt&\omit&&\omit&&\omit&&\omit&\cr
\noalign{\hrule}
height3pt&\omit&&\omit&&\omit&&\omit&\cr
&14&
&&&&&9&\cr
height3pt&\omit&&\omit&&\omit&&\omit&\cr
\noalign{\hrule}
height3pt&\omit&&\omit&&\omit&&\omit&\cr
&15&
&10&&&&15&\cr
height3pt&\omit&&\omit&&\omit&&\omit&\cr}
\hrule height 1.1pt}
$$
\vskip5pt
\noindent
{\bf Table 5.2b}: The spectrum for the massless matter fields
 in the $(2000)_{E_6}$ model.
There are 23 ${\bf 27}$'s, 7 ${\bf \bar {27}}$'s and 191 ${\bf 1}$'s of $E_6$
respectively. Note that when we consider the contributions from the
twisted sector $N$ mod 4 we get exact agreement with Table 5.2a, eg
there are 16 generations from $N=3$ mod 4.}

\medskip

\vbox{
$$\vbox{\offinterlineskip
\hrule height 1.1pt
\halign{&\vrule width 1.1pt#&\strut\quad#\hfil\quad&
\vrule#&\strut\quad#\hfil\quad&
\vrule#&\strut\quad#\hfil\quad&
\vrule#&\strut\quad#\hfil\quad&
\vrule#&\strut\quad#\hfil\quad&
\vrule#&\strut\quad#\hfil\quad&\vrule width 1.1pt#\cr
height3pt&\omit&&\omit&&\omit&&\omit&\cr
&\hfil$N$&&\hfil ${\bf 27}$&&\hfil ${\bf \bar{27}}$&&${\bf 1}$&\cr
height3pt&\omit&&\omit&&\omit&&\omit&\cr
\noalign{\hrule height 1.1pt\vskip3pt\hrule height 1.1pt}
height3pt&\omit&&\omit&&\omit&&\omit&\cr
&0&
&16&&&&52&\cr
height3pt&\omit&&\omit&&\omit&&\omit&\cr
\noalign{\hrule}
height3pt&\omit&&\omit&&\omit&&\omit&\cr
&1&
&6&&&&26&\cr
height3pt&\omit&&\omit&&\omit&&\omit&\cr
\noalign{\hrule}
height3pt&\omit&&\omit&&\omit&&\omit&\cr
&2&
&7&&2&&45&\cr
height3pt&\omit&&\omit&&\omit&&\omit&\cr
\noalign{\hrule}
height3pt&\omit&&\omit&&\omit&&\omit&\cr
&3&
&&&&&75&\cr
height3pt&\omit&&\omit&&\omit&&\omit&\cr
\noalign{\hrule}
height3pt&\omit&&\omit&&\omit&&\omit&\cr
&4&
&&&1&&10&\cr
height3pt&\omit&&\omit&&\omit&&\omit&\cr
\noalign{\hrule}
height3pt&\omit&&\omit&&\omit&&\omit&\cr
&5&
&&&2&&14&\cr
height3pt&\omit&&\omit&&\omit&&\omit&\cr
\noalign{\hrule}
height3pt&\omit&&\omit&&\omit&&\omit&\cr
&6&
&&&1&&9&\cr
height3pt&\omit&&\omit&&\omit&&\omit&\cr
\noalign{\hrule}
height3pt&\omit&&\omit&&\omit&&\omit&\cr
&7&
&1&&&&&\cr
height3pt&\omit&&\omit&&\omit&&\omit&\cr
\noalign{\hrule}
height3pt&\omit&&\omit&&\omit&&\omit&\cr
&8&
&&&2&&6&\cr
height3pt&\omit&&\omit&&\omit&&\omit&\cr
\noalign{\hrule}
height3pt&\omit&&\omit&&\omit&&\omit&\cr
&11&
&&&&&9&\cr
height3pt&\omit&&\omit&&\omit&&\omit&\cr
\noalign{\hrule}
height3pt&\omit&&\omit&&\omit&&\omit&\cr
&12&
&&&1&&&\cr
height3pt&\omit&&\omit&&\omit&&\omit&\cr
\noalign{\hrule}
height3pt&\omit&&\omit&&\omit&&\omit&\cr
&14&
&&&&&5&\cr
height3pt&\omit&&\omit&&\omit&&\omit&\cr
\noalign{\hrule}
height3pt&\omit&&\omit&&\omit&&\omit&\cr
&15&
&19&&&&&\cr
height3pt&\omit&&\omit&&\omit&&\omit&\cr}
\hrule height 1.1pt}
$$
\vskip5pt
\noindent
{\bf Table 5.2c}: The spectrum for the massless matter fields
 in the $(2200)_{E_6}$ model.
There are 49 ${\bf 27}$'s, 9 ${\bf \bar {27}}$'s and 251 ${\bf 1}$'s of $E_6$
respectively.}

\medskip

\vbox{
$$\vbox{\offinterlineskip
\hrule height 1.1pt
\halign{&\vrule width 1.1pt#&\strut\quad#\hfil\quad&
\vrule#&\strut\quad#\hfil\quad&
\vrule#&\strut\quad#\hfil\quad&
\vrule#&\strut\quad#\hfil\quad&
\vrule#&\strut\quad#\hfil\quad&
\vrule#&\strut\quad#\hfil\quad&\vrule width 1.1pt#\cr
height3pt&\omit&&\omit&&\omit&&\omit&\cr
&\hfil$N$&&\hfil ${\bf 27}$&&\hfil ${\bf \bar{27}}$&&${\bf 1}$&\cr
height3pt&\omit&&\omit&&\omit&&\omit&\cr
\noalign{\hrule height 1.1pt\vskip3pt\hrule height 1.1pt}
height3pt&\omit&&\omit&&\omit&&\omit&\cr
&0&
&&&&&452&\cr
height3pt&\omit&&\omit&&\omit&&\omit&\cr
\noalign{\hrule}
height3pt&\omit&&\omit&&\omit&&\omit&\cr
&2&
&&&&&35&\cr
height3pt&\omit&&\omit&&\omit&&\omit&\cr
\noalign{\hrule}
height3pt&\omit&&\omit&&\omit&&\omit&\cr
&4&
&&&&&16&\cr
height3pt&\omit&&\omit&&\omit&&\omit&\cr
\noalign{\hrule}
height3pt&\omit&&\omit&&\omit&&\omit&\cr
&5&
&&&1&&&\cr
height3pt&\omit&&\omit&&\omit&&\omit&\cr
\noalign{\hrule}
height3pt&\omit&&\omit&&\omit&&\omit&\cr
&15&
&149&&&&&\cr
height3pt&\omit&&\omit&&\omit&&\omit&\cr}
\hrule height 1.1pt}
$$
\vskip5pt
\noindent
{\bf Table 5.2d}: The spectrum for the massless matter fields
 in the $(2220)_{E_6}$  and $(2222)_{E_6}$ models.
There are 149 ${\bf 27}$'s, 1 ${\bf \bar {27}}$'s and 503 ${\bf 1}$'s of $E_6$
respectively.}

\newsec{Discussion and Outlook}
\nref\GVW{B. Greene, C. Vafa, and N. Warner, Nucl. Phys. {\bf B324} (1989)
371.}
\nref\Asymm{K. Narain, M. Sarmadi, and C. Vafa, Nucl. Phys. {\bf B288} (1987)
551.}
Our goal in this paper has been to describe how to use the (0,2) minimal
models (presented herein) as building blocks in constructing exactly soluble
(0,2) string
compactifications, much as Gepner did years ago for (2,2) models\gepner.
The correspondence between Gepner models and Calabi--Yau (or more precisely
Landau--Ginzburg) compactifications\GVW\ is an illuminating example of the
profound  connection between conformal field theory and spacetime geometry
in string theory.  It is therefore tempting to speculate about a similar
connection between exactly soluble (0,2) models and (0,2) Calabi--Yau
compactifications.  Indeed, the authors of ref.\cfttwo\ have provided
tantalizing hints of such a connection between their soluble models
and the (0,2) Landau--Ginzburg orbifolds described
in ref.\dkone.  It is our hope that detailed study of the class of models
described in this paper will yield further hints in this direction,
which may point the way towards a direct construction
connecting the
soluble models to specific Calabi-Yau compactifications as in ref.\GVW.

There are many interesting avenues of further study which present
themselves in this paper. One is the computation of the modular invariant
partition functions of the more general $G/H$ heterotic cosets, corresponding
to families of $(0,2)$ generalisation of the Kazama--Suzuki models. The closest
analogue among these to the $(0,2)$ minimal models studied in this paper would
be the case where  the symmetry $g\to gh$ is gauged, where $g\in G$ and $h\in
H$ and $G/H$ is Kahler. The modular invariant $(0,2)$ supersymmetric partition
function arising from this construction (after coupling in fermions and
canceling anomalies) would have states on the right coming from the $N=2$
Kazama--Suzuki series, assembled into the character ${\bar\X}^{N=2}$, and on
the left,
 there would be a $\X^G$ character, corresponding to the affine $G$-symmetry on
the left, together with states associated with  an $SO({\rm dim}G-{\rm
dim}H)/H$
coset, coming from the left--moving fermions. These models would again have
$c_L=c_R$, and it would be very interesting to study the spectrum of string
theories which can be constructed out of this.
Generically they would have gauge group $SO(10)$ together with a factor coming
from any affine symmetries present in the internal theory, as before. An
$E_6$ gauge group would again be realisable by using a left spectral flow
operator.

Another avenue of investigation is to compute the spectra of all of the
possible `doped' models which can be made  from tensor products of $(2,2)$
minimal models and at least one $(0,2)$ minimal model. Here, we have studied
only
the $(k=6)^4$ cases, (with and without $E_6$) which is only a small subset
of the possible $(0,2)$ exactly solvable
string vacua which can now be made using the
methods of this paper.

Of course, the focus in this paper has been  on $c_L=c_R$ compactifications.
It would be interesting to use  our methods to attack the problem of
finding exactly solvable $c_L>9$ $(0,2)$ vacua to complement those presented in
ref.\cfttwo.

Furthermore,  it is possible to extend even further the class of solvable
$(0,2)$ models by  considering extra  orbifolds,  as is
done for $(2,2)$ models.
This naturally led to the understanding of mirror symmetry via the
Greene--Plesser construction\ref\greenPless{B. R. Greene and
 M. R. Plesser, Nucl. Phys. {\bf B338} (1990) 15.}\ using  Gepner models.
 Work is in progress on whether such issues can be addressed
in the $(0,2)$ context\foot{Results akin  to this in the context of $(0,2)$
linear
sigma models have been presented in ref.\ref\jacquessham{J. Distler and S.
Kachru, Nucl. Phys. {\bf 442} (1995) 64, hep-th/9501111.}.}. Preliminary
results show that there are indeed
orbifold relations  (involving the aforementioned $\IZ_{2Q}$ symmetry)
analogous to those found for the $(2,2)$ models. Until
we have a better understanding of how the spectra we have computed relates
to a possible geometrical description, it is not yet
clear what the interpretation of such results will ultimately be.

This leads
us to the last (but not least) point.
There is the problem  of finding powerful arguments to
determine moduli among the gauge singlets in all of these models, sidestepping
the labour--intensive brute force calculation of all correlators directly from
the partition function. In the  examples presented, we were only able to
see moduli which preserved the enhanced gauge symmetries as well as the
$SO(10)$ or $E_6$ gauge group.
It would be of interest to find methods which can determine the moduli
which preserve only the generic $SO(10)$ or $E_6$ gauge group. These are
likely to be the generic gauge groups arising in the region of moduli space
connected  to a possible  sigma model geometrical
description. Such methods are going to become absolutely
indispensable if some understanding  of where these models lie in the moduli
space of $(0,2)$ compactifications is to be gained.

Of course, it is far from clear that one should expect the models
described herein (and others in this class)
to arise as special points in Calabi--Yau moduli
spaces.  Indeed, there are known examples of $N=1$ supersymmetric
compactifications where the internal
space is taken to be an asymmetric orbifold \Asymm\ which one does
not expect to be related by smooth deformation to a more conventional
geometry.  Thus is could equally well be that our models are describing
analogues of such asymmetric geometries, this time based on interacting
conformal field theories instead of free field theory.
These questions will be answered when we learn of more powerful ways of
determining the moduli of the models.

\bigskip

\noindent{\bf Note added}

After the appearance of this paper, we
were informed by Ralph Blumenhagen and Andreas Wisskirchen that
they have been able to reproduce the spectra of our $E_6$ examples using
their simple current program.

\vfill\eject

\noindent{\bf Acknowledgments}

The authors would like to thank Paul Aspinwall,   Matthias Blau,
Shyamoli Chaudhuri, Jacques Distler, Keith Dienes,
 Alon Farragi,
Mans Henningson, Wolfgang Lerche, Rob Myers,
Ronen Plesser, Rolf Schimmrigk, Eric Sharpe, Eva Silverstein,
Nick Warner  and  Edward Witten for helpful comments and
conversations.

The individual authors would like to thank each other for an extremely
enjoyable
collaboration over the last year or so. PB was supported by
funds provided by the U.S. Department of Energy (D.O.E.) under grant
 \# DOE--FG02--90ER40542 and would
like to acknowledge the hospitality of the Theory Division at CERN
and the Aspen Center for Physics. Some of this work was carried out while PB
was a member of the Institute for Advanced Study, Princeton.
CVJ was
supported by an EPSRC (UK) Postdoctoral
Fellowship, and would like to acknowledge the hospitality of the Trieste Centre
for Theoretical Physics. Some of this work was carried out while CVJ
 was a member of the Physics Department of  Princeton University.
PB and CVJ are currently supported in part by the
National Science Foundation under grant No. PHY94-07194.
 SK is supported by a fellowship from the Harvard Society of
Fellows and by the William F. Milton Fund of Harvard University.
PZ is supported in part by the Swiss National Science Foundation
and by funds provided by the U.S. Department of Energy (D.O.E.) under
cooperative agreement \# DF--FC02--94ER40818.
PZ would also like
to acknowledge the hospitality of the Institute for Advanced Study, Princeton.

\vfill\eject

\appendix{A}{{Elliptic Genus as a sum of $SU(2)$ affine characters}}
\noindent
\def\a{{\alpha}}
\def\X{{\cal X}}
We show here that
the elliptic genus \ellipticgenus\  is
\eqn\aone{
Z_{Q}(q,\gamma,0) = \sum_{r=0}^{Q-1} (-1)^r \X_{Q-1+2Qr;k}(\tau,z) ~,
}
where the complete affine $SU(2)$ characters at level $k$ are
\eqn\sutwoch{
\X_{l;k}(\tau,z) = {\Theta_{l+1;k} (\tau,z)- \Theta_{-l-1;k} (\tau,z) \over
\Theta_{1,2} (\tau,z)- \Theta_{-1,2} (\tau,z) }
= \sum_{m=0}^{2k-1} C^l_m(\tau) \Theta_{m;k}(\tau,z) ~,
} where the generalised $\Theta$--functions are defined in eqn.\thetagen.
The proof follows closely the one in ref.~\difran\ for a
similar situation in the (2,2) models.
The goal is achieved by first expressing the elliptic genus \ellipticgenus\
as a ratio of
Jacobi theta functions, then showing that the ratio `sum over characters'
divided by `elliptic genus' is an elliptic function and finally proving that
this ratio is a modular invariant function, actually equal to one.

We rewrite the elliptic genus using the first Jacobi theta
function
\ref\gsw{M. B. Green, J. Schwarz and E. Witten, {\sl `Superstring Theory'},
Cambridge University Press, Cambridge, 1987.}
\eqn\thetafunc{
\Theta_1(\nu|\tau) = [ 2 q^{1/4} \prod_1^\infty (1-q^{2n}) ]
\sin \pi \nu \prod_1^\infty (1-q^n \e{2 i\pi \nu})(1-q^n \e{- 2i \pi \nu}) ~.
}
If we put respectively
\eqn\respect{
\nu = {\gamma \over 2\pi} {1\over 2Q}, \qquad
\nu = {\gamma \over 2\pi} {1 \over k+2}
}
then the numerator and denominator of the elliptic genus are nothing
but theta functions (the square brackets cancel) and we get
\eqn\aeg{
Z_\a(q,\gamma,0) = {\Theta_1\left(
{\gamma \over 2\pi} {1\over 2Q} \mid \tau \right)  \over
\Theta_1 \left({\gamma \over 2\pi} {1 \over k+2} \mid \tau \right) }~.
}
We define the new variable
\eqn\newvars{
u={\gamma \over 2\pi} {1 \over k+2} = {\gamma \over 2\pi} {1 \over 2Q^2} ~,
}
and the elliptic genus \aeg\ becomes
\eqn\aegZ{
Z (\tau,u) = {\Theta_1(Qu|\tau) \over \Theta_1(u|\tau) } ~.
}
For later convenience, define the function $K(\tau,z)$
as the sum of characters
\eqn\adefK{
K(\tau,z) = \sum_{r=0}^{Q-1} (-1)^r \X_{Q-1+2Qr;k}(\tau,z) ~.
}

Now we look at the periodicity in the $u,z$ variables of the $Z,K$
functions. From standard properties of the Jacobi theta functions,
$Z$ behaves as
\eqn\behavesi{
\eqalign{
Z(\tau,u+2) &= Z(\tau,u) ~,\cr
Z(\tau,u+2\tau) &= \e{-4i\pi (Q^2-1)(\tau+u)} Z(\tau,u) ~.}
}
Similarly, due to the properties of the affine $SU(2)$ characters, $K$ behaves
as
\eqn\behavesii{\eqalign{
K(\tau,z+2) &= K(\tau,z) ~,\cr
K(\tau,z+2\tau) &= \e{-4i\pi (Q^2-1)(\tau+z)} K(\tau,z) ~.}
}
Therefore the ratio
\eqn\ratio{
R(\tau,z) = {K(\tau,2z) \over Z(\tau,2z)} = A \prod_i {\Theta_1(z-a_i|\tau)
\over \Theta_1(z-b_i|\tau) }}
is an elliptic function of $z$ (of periods 1 and $\tau$), and as such can
be expressed as a ratio of Jacobi
theta functions, where $a_i,b_i$ are respectively the
zeroes and poles of $R$ in the fundamental domain.

Next we show that this elliptic function $R(\tau,z)$ is modular
invariant. From ref.~\difran, we have
\eqn\aZmod{
\eqalign{
Z(\tau+1,2z) &= Z(\tau,2z) ~,\cr
Z(-1/\tau, -2z/\tau) & = \e{4 i \pi (Q^2-1) z^2/\tau} Z(\tau,2z) ~.}
}
Since the $SU(2)$
characters satisfy $\X_{n;k}(\tau+1,z) =\X_{n;k}(\tau,z)$, we get
\eqn\get{
K(\tau+1,2z) = K(\tau,2z)~.
}
For the $S$ transformation, recall that the $k+1$ characters span a linear
representation of the modular group
\eqn\aXtransf{
\X_{n;k}(-1/\tau,-z/\tau) = \e{i \pi k z^2/2\tau} \sqrt{{2 \over k+2}}
\sum_{n'=0}^k \sin \pi {(n+1)(n'+1) \over k+2} \;\; \X_{n';k}(\tau,z) ~.
}
We have to show that our particular sum of characters \adefK\
transforms into itself, up to an overall factor as in \aZmod.
After inserting
the proper values for $k=2(Q^2-1)$ and $n=Q-1+2Qr$ in \aXtransf,
one can prove that
\eqn\prove{
\sum_{r=0}^{Q-1} (-1)^r \sin \pi {(1+2r)(n'+1) \over 2Q} = Q
\sum_{r' \in \IZ}
(-1)^{r'} \delta_{n',Q-1+2Qr'}~.
}
So the $K$ transformation is
\eqn\kis{\eqalign{
&K(-1/\tau,-2z/\tau)  =\cr& \e{4 i \pi (Q^2-1) z^2/\tau} \sum_{n'=0}^{2(Q^2-1)}
\sum_{r' \in \IZ} (-1)^{r'} \delta_{n',Q-1+2Qr'} \; \X_{n';k}(\tau,2z) \cr
  &= \e{4 i \pi (Q^2-1) z^2/\tau} \sum_{r'=0}^{Q-1} (-1)^{r'}
\X_{Q-1+2Qr';k}(\tau,2z) \cr
 &= \e{4 i \pi (Q^2-1) z^2/\tau} K(\tau,2z) ~.}
}
This is the same factor as for the $Z$ function, and consequently $R(\tau,z)$
is a modular invariant.

The last step uses the lemma of ref.~\difran\ which states that an elliptic
function, which is also modular invariant, has to be a constant. We are left to
show that this constant is one, which is achieved by taking the
$q \rightarrow 0$ limit in both $Z,K$.

{}From the elliptic genus expression we get (putting $y=\exp(i\gamma/2Q^2)
= \exp(4i\pi z)$)
\eqn\fromellip{
Z(q=0,2z) = y^{-(Q-1)/2} {1-y^{Q} \over 1-y} =y^{-(Q-1)/2} \sum_{n=0}^{(Q-1)}
y^n ~,
}
and exactly the same result from the expansion of the $SU(2)$ characters
at $q=0$. Therefore the constant is one and the relation \aone\ is proven.

\appendix{B}{{Modular Invariance of $(0,2)$ models.}}
\noindent
In the partition function \intermpf,
the parameters at our disposal are $\tm,\tn$ but we also have a constraint on
$\bM$ coming from \intrm. Taking the $T$ invariance condition \tinv\ into
account, it is convenient then to introduce new unconstrained integers
$m, a_m, b$ related to the old variables by
\eqn\by{
{\tm \over Q} = {m \over 2}, \qquad
{\tn \over Q} = {m \over 2}+a_m, \qquad
\bar{M}= 4b - m(1+Q) - 2 a_m Q ~,
}
where for even $m,a_m \in 2\IZ$, and for odd $m, a_m \in \IZ$.
At this stage, this ensures the $T$ invariance of the partition function.
The problem now is the $S$--invariance of this object
\eqn\thisobj{
\eqalign{
  &Z(\tau) = \cr
&\sum_{m,a_m} \sum_{L,\bL=0}^{k} \sum_{b=0}^{Q^2-2}
\X_L (\tau) N_{L\bL}
\bC^\bL_{4b-m(1+Q)-2a_mQ} (\tau)
\bth_{(4b-m)Q^2-Q(m+2a_m); kQ^2} (\tau) \cr
  &= \sum_{L,\bL=0}^{k} \X_L (\tau) N_{L\bL} R_\bL(\tau) } }
which we will achieve by carefully choosing the sums over $m,a_m$.
(See~\sutwoch\ and \thetagen\ for the definition of the $SU(2)$ characters,
string functions and the generalized theta functions.)
Their range are determined by the various properties of the level $k$
string function
$C^l_m = C^l_{m+2k\IZ} = C^l_{-m} = C^{k-l}_{k-m}$ and theta function
$\Theta_{m+2k\IZ;k}=\Theta_{m;k}$.
Together with \thisobj, this implies that
$0 \leq m < 4Q(Q-1)$ and $0 \leq a_m < 2Q(Q^2-1)$.
Note also that the non vanishing of the string functions puts some restrictions
on the allowed spins $\bL \equiv m(1+Q) \,{\rm mod}\, 2$.
When $Q$ is odd, only even spins may occur.

Since the string functions and theta functions have a different level (and
hence different index periodicity), it is
convenient to rewrite
\eqn\rewrite{
m=m_0+m_1 4(Q-1), \qquad 0 \leq m_0 \leq 4(Q-1)-1,\quad 0 \leq m_1 \leq Q-1~,
}
and one sees that only the theta function index depends on $m_1$. Thus,
$a_m$ is constrained only by the value of $m_0$ and will be denoted $a_0$.
Next we
introduce the new variables
\eqn\uvamzero{
v = m_0+2a_0,
\qquad u = m_0(1+Q)+2 a_0Q = m_0+vQ ~,
}
which gives
\eqn\rrightfin{
R_{\bar L}(\tau) = \sum_{u,v,m_1} \sum_{b=0}^{Q^2-2}
\bC^{\bar L}_{4b-u} (\tau)
\bth_{(4b-u)Q^2 -(4m_1-v)Q(Q^2-1) ; kQ^2} (\tau)~. }
Both $u$ and $v$ can be defined modulo 4, due to the sum over $4b$, the
presence of $m_1$ and the
periodicity of the theta function. Note, however, that they
are not completely independent due to their very definition; we will return to
this point when we study the various cases more explicitly.

In this appendix, we will show that the following ansatz
\eqn\pfansatz{\eqalign{
  &Z(\tau) =\cr& \sum_{L,\bar{L}=0}^k \sum_{b=0}^{Q^2-2} \sum_{m_1=0}^{Q-1}
\sum_{u,v=0}^3
\X_L(\tau) N_{L\bar{L}}
\bar{C}^{\bar{L}}_{4b-u} (\tau)
\bth_{(4b-u)Q^2-(4m_1-v)Q(Q^2-1);kQ^2} (\tau)
}}
is modular invariant.
Note that $m_1$ is summed over its whole range.
Below we will specify the $u$ and $v$ sums,
which depend on the parity of $Q$, in order to ensure modular invariance
of $Z$.
Originally, this Ansatz was found by first performing a $S$ modular
transformation on \thisobj, then computing the $b$ sum and adjusting the
various sums in order to go back to the original expression \thisobj.

A first simplification in~\pfansatz\ occurs thanks to the identity
\eqn\trickth{
\sum_{a=0}^{Q-1} \Theta_{x-a 4Q(Q^2-1) ; kQ^2}(\tau,0) =
\Theta_{{x \over Q} ; k}(\tau,0) ~,
}
so that
\eqn\rrightfin{
  R_{\bar L}(\tau) = \sum_{u,v} \sum_{b=0}^{Q^2-2}
\bC^{\bar L}_{4b-u} (\tau) \bth_{(4b-u)Q + v(Q^2-1) ; k} (\tau)~.
}

For the next steps, we will need the transformation
properties under the $S$ modular transformation
of the $SU(2)$ characters $\X_{L;k}$, theta functions
$\Theta_{n;p}$ and string functions $C^L_{m;k}$ respectively,
\eqn\stransf{
\eqalign{
  \X_{L;k}({-1 \over \tau}) &= \sum_{L'=0}^k S_{LL'}\X_{L';k}(\tau)\,, \cr
  \Theta_{n;k}({-1 \over \tau}) &= \sqrt{-i\tau} \sum_{n'=0}^{2k-1}
s_{nn'} \Theta_{n';k}(\tau)\,, \cr
  C^L_{L+2m;k}({-1 \over \tau}) &= {1 \over \sqrt{-i\tau}} \sum_{L'=0}^k
\sum_{m'=0}^{k-1} S_{LL'} C^{L'}_{L'+2m';k} (\tau)
s^*_{L'+2m',L+2m} ~,\cr
}}
where
\eqn\wh{
S_{LL'} = \sqrt{{2\over k+2}} \sin \pi
{(L+1)(L'+1) \over k+2} , \qquad
s_{nn'} = {1 \over \sqrt{2k}} \e{-i\pi{nn' \over k}} ~.
}
Note that the $s$ matrix
depends on the level of the string or theta function.

Let us first study the $SU(2)$ modular invariance.
Recall that $C^L_M$ vanishes unless $L-M=0\,{\rm mod}\, 2$.
We can not in general
split the problem into its $SU(2)$ part ($L$ index) and its
$C\Theta$ part ($M$ index), since the $M$ index sum depends on the parity
of the $L$ index sum, as a consequence of~\stransf.
We will return to this as we study the individual
cases in more detail.
For the moment, it is useful to compute some partial sums on the $SU(2)$ side.

{}From~\ciz\ we have for
the $A$ and $D$ invariants,
\eqn\ADin{
\eqalign{
A_{k+1}\,:&\sum_{l=0}^k\X_l\bar\X_l ~,\cr
D_{{k\over 2}+2}\,:&\sum_{l=0}^{k/4-1}|\X_{2l}+\X_{k-2l}|^2+2|\X_{k/2}|^2~,
\qquad k\in 4\IZ\cr
D_{{k\over 2}+2}\,:&\sum_{l=0}^{k/2}|\X_{2l}|^2+
\sum_{l=0}^{k/2-1}\X_{2l+1}\bar\X_{k-2l-1}~,\qquad k\in 4\IZ + 2\cr
}}
We see that the $N_{L\bar L}$ are essentially of the form
$\delta_{L,\bL}$ or $\delta_{k-L,\bL}$. Applying~\stransf\ to~\pfansatz\ and
summing separately over $L=2l,2l+1$ we find (where $\epsilon=0,1$)
\eqn\sutwosum{
\eqalign{
  \sum_{l=0}^{k/2-\epsilon} S_{L',2l+\epsilon} & S^*_{2l+\epsilon,\bL'} =
{1 \over 2}(\delta_{L',\bL'}+(-1)^\epsilon \delta_{k-L',\bL'}) ~,\cr
  \sum_{l=0}^{k/2-\epsilon} S_{L',2l+\epsilon} & S^*_{k-2l-\epsilon,\bL'} =
{(-1)^{L'} \over 2}(\delta_{L',\bL'}+(-1)^\epsilon\delta_{k-L',\bL'}) ~.\cr
}}

The next step is to study the behaviour of~\rrightfin\ under $S$, looking
at the $M$ index part only
\eqn\sright{\eqalign{
&R_{\bar L'}({-1\over\tau}) =\cr & {1\over 2kQ}
\sum_{u,v} \sum_{b=0}^{Q^2-2} \sum_{M',N'=0}^{2k-1}
\e{-{i\pi M'\over k}(4bQ^2-u)}
\e{{i\pi N'\over k}((4b-u)Q+v(Q^2-1))}
C^{\bL'}_{M'}(\tau) \Theta_{N';k}(\tau)~.
}}
Note that we have used that $C_{4b}=C_{4bQ^2}$ due to the periodicity
$2k=4(Q^2-1)$ of $C$. We first perform the sum over $b$,
\eqn\bsum{
\sum_{b=0}^{Q^2-2} \e{-{i\pi 4b\over k}(M'Q^2-N'Q)}=
(Q^2-1) \sum_{c=0}^{4Q^2-1} \delta_{N',M'Q+c(Q^2-1)/Q}~.
}
Since $Q^2-1$ and $Q$ are relatively prime, it must be that $c=\bar c Q$.
Next, by rewriting $\bar c$ as
\eqn\rew{
\bc=\bc_0 + 4 \bc_1, \qquad 0 \leq \bc_0 \leq 3, \quad 0 \leq \bc_1 \leq Q-1~,
}
neither the phase factor in~\sright\ nor the theta function depend on
$\bar c_1$, and performing the
sum over $\bar c_1$
just yields a multiplicative factor of $Q$.
We obtain
\eqn\srightfin{\eqalign{
R_{\bar L'}({-1 \over \tau})
= {1 \over 4} \sum_{u,v} \sum_{M'=0}^{2k-1} \sum_{\bc_0=0}^{3}&\biggl\{
\bC^{\bL'}_{M'} (\tau)
\bth_{M'Q+\bc_0(Q^2-1);k}(\tau) \,\times\cr
&\e{i \pi {v \over 2}(M'Q+\bc_0(Q^2-1))}
\e{-i \pi {u \over 2}(M'+\bc_0 Q)} \biggr\}\,,
}}
Note the close resemblance between~\srightfin\ and~\rrightfin. In order to
continue we need to specify the type of modular $SU(2)$ modular invariant
which is used as well as the parity of $Q$.

\subsec{$Q$ odd, non--diagonal invariant}
This particular $SU(2)$ invariant, $D_{{k\over 2}+2}$,
involves only even spins, and there is only the even sum in \ADin.
Furthermore, the sum over even spins only is enough to have $SU(2)$ invariance,
\eqn\evenspins{
\sum_{l,\bar{l}=0}^{k/2} S_{L',2l} D_{2l,2\bar{l}} S^*_{2\bar{l},\bL'}
= D_{L',\bL'}~.
}
Since the left-moving $SU(2)$ part was invariant by itself, we just have to
show that the right-moving
$C\Theta$ part is also invariant, independently of the value of $\bL'$.

First we should see what the ranges for $u,v$ are; this depends on $Q$
but not on the $SU(2)$ modular invariant chosen. For $Q$ odd choose the
following two possibilities
\eqn\uvqodd{
\left\{ \eqalign{
&0 \leq v \leq 3 \cr
& u=0} \qquad{\rm and} \right. \qquad
\left\{ \eqalign{&v=0,2 \cr &u=0 }  \right. \quad
\left\{ \eqalign{&v=1,3 \cr &u=2 } \right.
}
Of course this is in agreement with the restrictions on $u,v$ from $m_0,a_0$,
see~\uvamzero.

With the first choice in~\uvqodd, we have only the sum over $v$ in~\srightfin\
(for $Q$ odd, $Q^2-1 \in 8\IZ$ and the phase is independent of $\bc_0$)
\eqn\phase{
\sum_{v=0}^3 \e{i\pi{v \over 2}M'Q} = 4 \delta_{M'Q,4\IZ} =
4 \delta_{M',4\IZ}
}
since $Q$ is odd.
Therefore we get from~\srightfin\
\eqn\from{
R_{\bL'} ({-1 \over \tau}) =
 \sum_{m'=0}^{k/2-1} \sum_{\bc_0=0}^{3} \bC^{\bL'}_{4m'} (\tau)
\bth_{4m'Q+\bc_0(Q^2-1);k}(\tau)
}
which is exactly~\rrightfin\ with this particular choice of $u,v$.

For the second choice in \uvqodd,
we do separately the sums over even and odd $v$. One
finds that $M=2m'$ with $m'$ and $\bc_0$ simultaneously odd or even.
%$$
%\sum_{u=0;v=0,2} \e{i\pi{v \over 2}M'Q} = 2 \delta_{M'Q,2\IZ} =
%2 \delta_{M',2\IZ} .
%$$
%and
%$$
%\sum_{u=2;v=1,3} \e{i\pi{v \over 2}M'Q} \e{-i\pi(M' + \bc_0 Q)}
%= 2 \e{i\pi{1 \over 2}M'Q} (-1)^{M' + \bc_0 }
%\delta_{M'Q,2\IZ} = 2
%(-1)^{{M'\over 2} + \bc_0 }\delta_{M',2\IZ}
%$$
%since $M'$ is even and $Q$ is odd.
%Combining these two expressions with the 1/4, we get the projector
%$$
%{1+(-1)^{m' + \bc_0} \over 2} \delta_{M',2m'}
%$\eqno{(15)}
%$$
%which implies that both $m'$ and $\bc_0$ are simultaneously odd or even.
Then we get from~\srightfin\
\eqn\weget{
\eqalign{
  R_{\bL'} ({-1 \over \tau}) &=
 \sum_{m'=0}^{k/2-1} \sum_{\bc_0=0,2} \bC^{\bL'}_{4m'}(\tau)
\bth_{4m'Q+\bc_0(Q^2-1);k}(\tau)  \cr
  &+ \sum_{m'=0}^{k/2-1} \sum_{\bc_0=1,3} \bC^{\bL'}_{4m'-2}(\tau)
\bth_{(4m'-2)Q+\bc_0(Q^2-1);k}(\tau) }
}
which is the same as~\rrightfin\ for this second choice of $u,v$ in~\uvqodd.

\subsec{$Q$ odd, diagonal invariant}
As above the constraint $\bL \equiv m(1+Q)\,{\rm mod}\, 2$ eliminates odd
spins.
The major difference now is that the sum over even spins is not enough to
transform the diagonal invariant into itself. Instead, using~\sutwosum\
we get
\eqn\srightdiag{
Z({-1 \over \tau}) =
\sum_{\bL',L'=0}^k\X_{L'}(\tau) \half (\delta_{\bL',L'} + \delta_{\bL',k-L'})
R_{\bar L'}({-1\over\tau})~.}
%Now, the sums over $L',\bL'$ run over all the integers, so it seems that we
%get also the odd spins after the $S$ transformation. But the $C\Theta$ part
%will get rid of the odd spins.

For each value of $\bL'$ we have on the right--moving side exactly the
same expression as~\rrightfin. The first choice in~\uvqodd\
implies that $M'=\bL'+2m' \in 4\IZ$, which implies
also that $\bL'$ is even, hence there is no odd spins! We get thus
\eqn\polkadots{\eqalign{
Z({-1 \over \tau}) = \sum_{l=0}^{k/4}\biggl\{
\half \X_{2l} (\tau) &\bC^{2l}_{4b} (\tau) \bth_{4bQ+\bc(Q^2-1);k}(\tau)\cr
&+ \half \X_{2l}(\tau) \bC^{k-2l}_{4b}(\tau) \bth_{4bQ+\bc(Q^2-1);k}(\tau)
\biggr\}
}}
where $\bc$ runs from 0 to 3.
The first term is fine, but we must work on the second term,
to remove the $k-2l$. We use the
symmetry of the string function $C^l_m = C^{k-l}_{k+m}$, the periodicity of
the theta function and the fact that
$k=2(Q^2-1) \in 16\IZ \subset 4\IZ$ to show that
\eqn\codd{
%\eqalign{
  \bar C^{k-2l}_{4b}(\tau) \bar\Theta_{4bQ+\bc(Q^2-1);k} (\tau)
%C^{2l}_{4b+k} \Theta_{4bQ+\bc(Q^2-1);k} \cr
%  &= C^{2l}_{4b'} \Theta_{(4b'-k)Q+\bc(Q^2-1);k} \cr
%  &= C^{2l}_{4b'} \Theta_{4b'Q+\bc(Q^2-1)-2(2q+1)(Q^2-1);k} \cr
  = \bar C^{2l}_{4b'}(\tau) \bar\Theta_{4b'Q+\bar c'(Q^2-1);k}(\tau)
}
where $4b'=4b+k$ and $\bar c'=\bc-2$.
Since we are summing over all values of $b$ allowed by the periodicity of the
string and theta functions, we can absorb the shift by $k$. The same is true
for the shift in $\bc$, since we sum over all allowed values for $\bc$.
Therefore
\eqn\swap{
\Sb \sum_{\bc=0}^3 \bar C^{k-2l}_{4b}(\tau) \bar\Theta_{4bQ+\bc(Q^2-1);k}(\tau)
=\Sb \sum_{\bc=0}^3 \bar C^{2l}_{4b}(\tau) \bar\Theta_{4bQ+\bc(Q^2-1);k}(\tau)
}
and the modular invariance of \rrightfin\ for the first choice in~\uvqodd\
has been proven.
%$$
%Z({-1 \over \tau}) = \X_{2l} \bC^{2l}_{4b} \bth_{4bQ+\bc(Q^2-1);k}(\tau)
%=Z(\tau)
%$$

For the second choice in~\uvqodd, we first do the sums over even and odd $v$
separately, as for the non-diagonal $SU(2)$ invariant. As above, we can then
rewrite the $\bar C^{k-2l}_{..}$ as $\bar C^{2l}_{..}$ which finally leads
to the
invariance
%
%For the second choice, the sum over $u,v$ introduces the projection (15), and
%we get
%$$
%\eqalign{
%  Z({-1 \over \tau}) &= \sum_{m'}
%\X_{L'} \half(\delta_{L',\bL'} + \delta_{k-L',\bL'}) \bC^{\bL'}_{2m'}
%\; \half(1+(-1)^{m'+\bc}) \; \bth_{2m'Q+\bc(Q^2-1);k}(\tau) \cr
%  &= \half \sum_{\bc=0,2} \Sb  \X_{2l} \bC^{2l}_{4b}
%\bth_{4bQ+\bc(Q^2-1);k} + \X_{2l} \bC^{k-2l}_{4b}
%\bth_{4bQ+\bc(Q^2-1);k} \cr
%  &+ \half \sum_{\bc=1,3} \Sb \X_{2l} \bC^{2l}_{4b-2}
%\bth_{(4b-2)Q+\bc(Q^2-1);k} + \X_{2l} \bC^{k-2l}_{4b-2}
%\bth_{(4b-2)Q+\bc(Q^2-1);k} }
%$$
%Again, we can use the same trick (19) with the symmetry of the $C$. Since $k
%\in 4\IZ$, then $k+4b-2 \in 4\IZ+2$, so we can replace it by $4b'-2$.
%Although the sum over $\bc$ is split now, the
%shift of $\bc$ in (18) is even so it can still be absorbed by a redefinition
%of $\bc$; therefore the two $k-2l$ contributions above are equal to their $2l$
%partners and we get modular invariance
\eqn\moonbeams{
\eqalign{
Z({-1 \over \tau}) = &\sum_{\bc=0,2} \X_{2l}(\tau) \bC^{2l}_{4b}(\tau)
\bth_{4bQ+\bc(Q^2-1);k}(\tau) \cr
+&\sum_{\bc=1,3} \X_{2l}(\tau)
\bC^{2l}_{4b-2}(\tau)
\bth_{(4b-2)Q+\bc(Q^2-1);k}(\tau)  = Z(\tau)~.}
}

\subsec{$Q$ even, diagonal invariant}
We start by reexamining the allowed values for $u,v$. Carefully
looking at~\uvamzero\ we find at
least two allowed possibilities which can be expressed
compactly as $u=\pm v$.

In comparison with the earlier cases, odd $Q$, there are two
important differences. First, we have to split the sum over $L$ (depending
on its parity) in taking
the $S$ transformation, which yields, after solving the $SU(2)$ part with the
help of~\sutwosum\
\eqn\pfsevendiag{
\eqalign{
  Z({-1 \over \tau}) &= \sum_{\bL',L'=0}^k
\X_{L'}(\tau) \half(\delta_{L',\bL'}+\delta_{k-L',\bL'})
R^{even}_{\bL'}({-1\over\tau})\cr
  &+ \sum_{\bL',L'=0}^k
\X_{L'} (\tau) \half(\delta_{L',\bL'}-\delta_{k-L',\bL'})
R^{odd}_{\bL'}({-1\over\tau})~.
}}
The notation $R^{even}$ ($R^{odd}$) means that this term contains only a sum
over even (odd) values of $u=\pm v$, as it corresponds to even (odd) spin
sums, and $u \equiv \bL\,{\rm mod}\,2$ from the non vanishing of the string
function in \rrightfin.
Second, the phase factor in~\srightfin\ can be
simplified, using the fact that $Q^2 \in 4\IZ$.

We can immediately add up the first and the third term together, and the
$v$ sums complement each other to give ($\bc_0$ is $\bc$ now)
\eqn\vsum{
\sum_{v=0}^3 \e{i \pi {v \over 2}(M'(Q \mp 1)-\bc(1 \pm Q))}
= 4 \delta_{M'(Q \mp 1)-\bc(1 \pm Q), 4\IZ}
= 4 \delta_{M',4\IZ \pm \bc}~.
}
Therefore we get a first contribution to~\pfsevendiag
%One difference will be in the phases summed over $u,v$ in~\srightfin\
%$$
%\e{i \pi {v \over 2}(M'Q+\bc(Q^2-1))}
%\e{-i \pi {\pm v \over 2}(M'+\bc Q)} =
%\e{i \pi {v \over 2}(M'(Q \mp 1)-\bc(1 \pm Q))}
%$$
%where we discarded $Q^2 \in 4\IZ$. The sum gives
%We want to prove the invariance of
%$$
%Z(\tau) = \Sb \sum_{v=0}^3 \X_L \bC^L_{4b \mp v} \bth_{(4b \mp v)Q +
%v(Q^2-1);k}(\tau) .
%$$
\eqn\pfhalf{
Z({-1 \over \tau})^{LL}= \half \sum_{L=0}^k \sum_{\bc=0}^3
\X_L (\tau) \bC^L_{4b \mp \bc} (\tau)
\bth_{({4b \mp \bc})Q + \bc(Q^2-1);k}(\tau)~.
}

For the two $\delta_{k-L',\bL}$ terms, there is a relative minus sign, which
can be expressed as $(-1)^v$, so the only change we have to do in recombining
these terms is to insert this sign in~\vsum. This gives
\eqn\thisgives{
%\sum_{v=0}^3 (-1)^v \e{i \pi {v \over 2}(M'(Q \mp 1)-\bc(1 \pm Q))} =
\sum_{v=0}^3 \e{i \pi {v \over 2}(M'(Q \mp 1)-\bc(1 \pm Q)+2)}
%= 4 \delta_{M'(Q \mp 1)-\bc(1 \pm Q), 4\IZ+2}
= 4 \delta_{M',4\IZ +2 \pm \bc}
}
and a second contribution to~\pfsevendiag
\eqn\whoa{
Z({-1 \over \tau})^{L,k-L}= \half
\sum_{L=0}^k \sum_{\bc=0}^3
\X_L (\tau) \bC^{k-L}_{4b +2 \mp \bc} (\tau)
\bth_{({4b +2 \mp \bc})Q + \bc(Q^2-1);k}(\tau) ~.
}

Now we use the symmetry of $C$ as in~\codd\ with the difference that
$k=2(Q^2-1) \in 8\IZ-2 \subset 4\IZ+2$, so that
$4b+2+k \in 4\IZ$, and also $Qk \in 2k\IZ$, and
\eqn\ceven{
\eqalign{
  \bar C^{k-L}_{4b+2 \mp \bc} (\tau)
\bar\Theta_{(4b+2\mp \bc)Q+\bc(Q^2-1);k} (\tau)
%C^{L}_{4b+2\mp \bc+k} \Theta_{(4b+2\mp \bc)Q+\bc(Q^2-1);k} \cr
%  &= C^{L}_{4b' \mp \bc} \Theta_{(4b'-k)Q+\bc(Q^2-1);k} \cr
%  &= C^{L}_{4b' \mp \bc} \Theta_{4b'Q+\bc(Q^2-1)-Qk;k} \cr
  = \bar C^{L}_{4b' \mp \bc}(\tau) \bar\Theta_{4b'Q+\bc(Q^2-1);k}(\tau)~. }
}
Thus,
\eqn\thusly{\eqalign{
\Sb \sum_{\bc=0}^3   \bar C^{k-L}_{4b+2 \mp \bc}(\tau)
&\bar\Theta_{(4b+2\mp \bc)Q+\bc(Q^2-1);k} (\tau)=\cr&
\Sb \sum_{\bc=0}^3 \bar C^{L}_{4b \mp \bc}(\tau)
\bar\Theta_{4bQ+\bc(Q^2-1);k}(\tau)
}}
which combines with~\pfhalf\ to give us~\invpm.

\subsec{ $Q$ even, non--diagonal invariant}
{}From~\ADin\ this $SU(2)$ invariant naturally splits into
odd and even spins and is
\eqn\butofcourse{\eqalign{
Z(\tau) = \Sb \sum_{v=0}^3 \biggl\{\X_{2l} \bC^{2l}_{4b \mp v} &\bth_{(4b \mp
v)Q +
v(Q^2-1);k}(\tau) \cr&+ \X_{2l+1} \bC^{k-2l-1}_{4b \mp v} \bth_{(4b \mp v)Q +
v(Q^2-1);k}(\tau) ~.\biggr\}
}}
The even sum is the same as in the diagonal case, but the odd sum has an extra
overall $(-1)^{L'}$ due to the $k-2l-1$ in $C$ (compare
with~\pfsevendiag).
Therefore, when $L'$ is even
there is no difference with the diagonal case and summing both the $LL$ and
$L,k-L$ contributions yields
\eqn\yieldknave{
Z^{even}({-1 \over \tau})
= \Sb \sum_{\bc=0}^3 \X_{2l}(\tau) \bC^{2l}_{4b \mp \bc}(\tau)
\bth_{(4b \mp \bc)Q + \bc(Q^2-1);k}(\tau)
}
which is fine.

For the odd $L'$ spin, the extra $-1$ kills the term from $\delta_{k-L',\bL'}$,
and the two contributions add up with no relative sign, giving
\eqn\therelatives{
Z^{odd}({-1 \over \tau})
= \Sb \sum_{\bc=0}^3 \X_{2l+1}(\tau) \bC^{k-2l-1}_{4b \mp \bc}(\tau)
\bth_{(4b \mp \bc)Q + \bc(Q^2-1);k}(\tau)
}
which is also fine. Here we chose to express the $\bar C^{2l+1}$ in terms of
$\bar C^{k-2l-1}$ in order to match the original expression.

\appendix{C}{The Elliptic Genus and the $(0,2)$ Modular Invariants}

We show that the elliptic genus \ellipic\ is proportional to
$Z'=Z_+-Z_-$. Taking the $SU(2)$ diagonal modular invariant for
simplicity, it is suggestive to write it as
\eqn\sugg{\eqalign{
 Z'= \sum_{odd~L} \X_L N_{L \bar L} \sum_{m=0}^{Q-1} \biggl\{&\left(
\bar \X^{L}_{(4m+1)Q;Q-1}-\bar \X^{L}_{(4m+1)Q;Q-3}\right)\cr
&+\left(\bar \X^{L}_{(4m+3)Q;Q-3}-\bar \X^{L}_{(4m+3)Q;Q-1}
\right)\biggr\}.
}}
Since the quantities in brackets have the same $L,q$ values but $s$
different by 2, they represent
${\rm Tr}(-1)^F q^{L_0 -c/24}$ taken in the {\sl complete} $N=2$
representation. Due to the unbroken super symmetry in the Ramond sector,
this is non zero only for those representations containing a Ramond
ground state. In the parametrisation where $0 \leq l \leq k$, $-k-1 \leq q
\leq k+2$, $-1 \leq s \leq 2$, the ground states appear in $\X^l_{l+1;1}$ or
$\X^l_{-l-1;-1}$. We need to find for which values of $l$ these characters
occur in \sugg.

Consider first $0 \leq m < Q/2$ (so that $0 < q < k+2$). In that case,
we must look for $\X^l_{l+1;1}$. If $Q-1 \equiv 1$, it occurs for
$l=(4m+1)Q-1$ with a plus sign, and for $l=(4m+3)Q-1$ with a minus
sign. For $Q-3 \equiv 1$, the same values of $l$ are selected, but the
signs are reversed. In short, we get the values
\eqn\lval{
\sum_{n=0}^{Q-1}
\delta_{l,(2n+1)Q-1} (-1)^{n+{Q \over 2}+1}
}

For the other range $Q/2 \leq m \leq Q-1$, we should subtract
$2(k+2)=4Q^2$ to $q$ for it to fall in the proper domain, and we must
look for the occurrence of $\X^l_{-l-1;-1}$. When $Q-3 \equiv -1$,
this happens for $l=(4(Q-m-1)+3)Q-1$ with a minus sign, and for
$l=(4(Q-m-1)+1)Q-1$ with a plus sign. When $Q-1 \equiv -1$, the signs
are reversed. This range gives the contribution
\eqn\lvall{
\sum_{n=0}^{Q-1}
\delta_{l,(2n+1)Q-1} (-1)^{n+{Q \over 2}+1}
}

Combining the two, we find that the result is
\eqn\humm{
Z_+ - Z_- = 2 (-1)^{{Q \over 2}+1}\sum_{n=0}^{Q-1}(-1)^{n}
\X^{su(2)}_{(2n+1)Q-1;k} = 2 (-1)^{{Q \over 2}+1} Z_Q^{(0,2)}(q,0,0) .
}
Since the set of allowed values for $l$ is symmetric under $l \to k-l$,
this proof is also valid for the other types of $SU(2)$ modular invariants.

A proof along the same lines is valid for the $Q$ odd case.

\listrefs
\bye